\documentclass[10pt,journal,compsoc]{IEEEtran}

\usepackage{authblk}
\usepackage{amsmath}
\usepackage[utf8]{inputenc}
\usepackage[table,xcdraw]{xcolor}
\usepackage[normalem]{ulem}
\usepackage{rotating}\usepackage{amsmath,amssymb}
\usepackage[linesnumbered,ruled,vlined]{algorithm2e}
\usepackage{changepage}
\usepackage{tcolorbox}
\usepackage{makecell}
\usepackage{subfigure}
\usepackage[colorlinks]{hyperref}
\usepackage[nameinlink,noabbrev]{cleveref}
\usepackage{booktabs}
\usepackage{graphicx}
\usepackage[square,sort,comma,numbers]{natbib}
\usepackage[thinlines]{easytable}
\newcommand{\tabincell}[2]{\begin{tabular}{@{}#1@{}}#2\end{tabular}}  

\usepackage{float}
\usepackage{diagbox}
\usepackage{stfloats}
\usepackage{multirow}

\newcommand{\killme}[1]{}

\definecolor{table_gray}{gray}{.9}

\begin{document}

\title{\textbf{DAMIA}: Leveraging \textbf{D}omain \textbf{A}daptation as a Defense against \textbf{M}embership \textbf{I}nference \textbf{A}ttacks }

%
\author{
    Hongwei Huang \thanks{Hongwei Huang,Weiqi Luo,Guoqiang Zeng, Jian Weng, Yue Zhang, Anjia Yang are with the College of Informatin Science and Technology / College of Cyber security, National Joint Engineering Research Center of Network Security Detection and Protection Technology, and Guangdong Key Laboratory of Data Security and Privacy Preserving, Jinan University, Guangzhou, 510632, China;}, 
    Weiqi Luo \thanks{Weiqi Luo is the corresponding author : lwq@jnu.edu.cn ;}, 
    Guoqiang Zeng, 
    Jian Weng, 
    Yue Zhang, 
    Anjia Yang

}

\IEEEtitleabstractindextext{%
  
\begin{abstract}


Deep Learning (DL) techniques allow one to train a model from a dataset, and the model can be adopted to solve tasks.
DL has attracted much interest given its fancy performance and potential market value, while security issues are amongst the most colossal concerns. Being a core engine of DL, models may be prone to the membership inference attack, where an attacker determines whether a given sample is from the dataset the model trained on.  Efforts have been made to hinder the attack but unfortunately, they may be subject to a major overhead or impaired usability. In this paper, we propose and implement DAMIA,  leveraging Domain Adaptation (DA) as a defense to counter membership inference attacks. Our observation is that during the training process, DA obfuscates the dataset to be protected using another related dataset, and derives a model that underlyingly extracts the features from both the original dataset and the introduced dataset. Seeing that the model is obfuscated, membership inference fails, while the extracted common features provide supports for usability. Extensive experiments have been conducted, which validates our intuition. The model counters the membership inference attacks and has a negligible footprint to the usability. Our experiment also excludes a few factors that may hinder the performance of DAMIA, providing a potential guideline that instructs the vendors and researchers to benefit from our solution in a timely manner.

\end{abstract}

}
\maketitle

\section{Introduction}
\vspace{-1mm}

Deep Learning (DL) is a subfield of machine learning, and it is inspired by the working of human brains in data processing.
Specifically, DL forms a mathematical model based on sample data, i.e., the training data, and progressively extracts higher level features from sample data, based on which the model can make decisions without human's involvement. 
Due to the fancy performance, DL has been widely adopted in a large range of domains, including image classification \cite{oord2016wavenet, mnih2015human} object recognition \cite{parkhi2015deep}, person re-identification \cite{li2014deepreid}, and disease prediction \cite{mohanty2016using}. As an illustration of such a trend, the market of DL is booming and estimated to hit USD 7.2 billion during 2020-2024, according to the statistics from Technavio \cite{technavio2020}.

While DL is penetrating the academia and industry, its explosive growth and huge potential also attract cybercriminals, bringing the rampant security issues against the DL community. 
In general, the model may be publicly accessible, while training data, as well as the properties of training data, are considered confidential. Therefore, extracting the training samples and the related information via the model is a violation of the security setting in DL, which has been widely discussed in previous efforts \cite{fredrikson2015model,shokri2017membership, ganju2018property, salem2019updates}. 
Among the attacks, the membership inference attack, which was proposed by Shokri et al. \cite{shokri2017membership}, has attracted a lot of recent attention \cite{pyrgelis2017knock,chen2019gan,liu2019socinf}. In this attack, the attacker may craft a malicious inference model based on predictions of a victim model. Due to the fact that a model has a better performance when a sample is from the original training dataset, the attacker may use the inference model to determine if a sample is from the training dataset of a victim model.

A model without any capability against membership inference attacks may lead to grave consequences, in that the DL models are adopted on large-dimension (and potentially sensitive) user dataset. For example, \citet{pyrgelis2017knock} present that membership inference is feasible on aggregate location data in real-world settings, where the attacker can reveals a user’s locations and actively traces the user. \citet{chen2019gan} deploy membership attacks against medical records, and the involved patients may be plagued by disease discrimination issues. Further, training data leakage may spawn pecuniary losses or even legal disputes for enterprises. According to GDPR (General Data Protection Regulation) \cite{voigt2017eu}, vendors now are forced to protect user privacy, and violations of GDPR may be imposed a penalty up to 20 million euros, or 4\% of the offender's global turnover of the preceding fiscal year.

Efforts have been made to counter membership inference attacks, while they are plagued by either heavily overhead or hindered usability.  These efforts can be categorized into three groups: ($i$) regularization-based defenses; ($ii$) adversarial-attack based defenses; ($iii$) differential-privacy-based defenses. Regularization-based defenses adapt regularization techniques to design countermeasures; however, it may lead to heavy overhead when complex regularization techniques are adapted. For example, \citet{salem2018ml} demonstrate that ensemble learning (a regularization technique) can be adapted to counter against membership inference attack, but multiple types of ML models are required, which brings significant training and storage costs. Adversarial-attack-based defenses utilize adversarial examples \cite{goodfellow2014explaining} to obfuscate the membership inference attack model, but the defenses require extra efforts in finding proper perturbations. For instance, \citet{jia2019memguard} propose a defense called MemGuard to modify the outputs of the victim model into adversarial examples, but a model mimicking the attacker is required to assist finding perturbations, which is time consuming. Differential-Privacy-based defenses adopt differential privacy to add noises during model training to thwart the attacks. However, the usability of the model is also impaired as discussed in \cite{pyrgelis2017knock, jayaraman2019evaluating}.


 It can be observed from these previous works \cite{pyrgelis2017knock, jayaraman2019evaluating} that to counter against membership inference attack,  one conventional wisdom is to introduce perturbations or noises. However, those newly introduced perturbations/noises also downgrade the usability of the model. Therefore, exploring a solution that may bridge the gap between usability and security is a vital challenge. In this paper, we leverage Domain Adaptation (DA) to build a defense in defending membership inference attacks.  DA allows knowledge to be transferred from a source domain to a different (but related) target domain. For example, DA may utilize images dataset of cats and dogs collected from Instagram (i.e., source domain) to solve new tasks such as categorizing pictures of cats and dogs clipped from animation movies (i.e., target domain). To this end, DA may train a shared representation (in our context, a shared representation is learned by a model) from the dataset in the source domain and the dataset in the target domain, and the shared representation shares the underlying common features of the two datasets. 
Our observation is that the two datasets are mixed and obfuscated when DA is adopted. 
Intuitively, if the sensitive dataset is in the target domain, we can find a different but related dataset in a source domain, and leverage DA to obfuscate the sensitive dataset. In such a way, membership inference attacks fail while the newly generated shared representation may also have a good performance in solving tasks requiring a sensitive dataset since the shared representation contains the features from the sensitive dataset. Also, the overhead of our defense is slight to none, in that there are no extra phases/algorithms that are involved in the model once it has released. 



To validate our intuition, we design and implement DAMIA \footnote{Damia is a Greek goddess who brings the fertility of the soil, and we hope our DAMIA also brings the ``fertility'' of DL community by mitigating the membership inference attacks. }, leveraging Domain Adaptation (DA) as a defense to counter membership inference attacks. We conduct extensive experiments to benchmark a balance where membership inference attacks are defended while the usability of the model is not affected. In terms of that, multiple metrics have been evaluated, including the effectiveness, accuracy and various other metrics. Of all the metrics, we also design a few novel metrics, which greatly outlines the capability of DAMIA. According to our experiments, while settings may vary, our defense always has a good performance in defending the membership inference attacks. Specifically, the success rate of an attacker is close to 50\%,  which is roughly equivalent to a random guess.  However, with higher similarity between the source dataset and the sensitive dataset, the accuracy of the model (i.e., the usability) shall be significantly boosted. Given that, we argue that our defense will not hinder the usability of the original model and has negligible fingerprints.

 {\textbf{Contributions}} Our paper makes the following contributions:
\begin{itemize}
    \item  We propose that DA is feasible to be leveraged against membership inference attacks. To our best known, we are the first to adopt DA in the domain of defending membership inference attacks.
    
    \item We design and implement DAMIA (i.e., Domain Adaptation against Membership Inference Attacks). Our experiment shows that DAMIA is capable of defending against membership inference attacks with high performance up to 50\%, which is roughly equivalent to a random guess. 
    
    \item We design a few metrics that can measure a model's capability of defending against membership attacks. We believe some of the metrics can better reflect the capability when compared with the legacy ones. 
    
    \item We further investigate a few factors that may affect our results in terms of accuracy. Our attempts may also bring benefits and values for vendors and researchers that have interests in adopting our defense, in that we excluded a few factors that may impair the usability of our defense.

\end{itemize}

\textbf{Roadmap}:
The rest of the paper is organized as follows. 
In \Cref{sec:background}, we provide background on membership inference attacks and domain adaptation.
In \Cref{sec:leveraging_domain_adaptation_as_a_defense} we define the threat model and represent the insight and the design of DAMIA.
In \Cref{sec:evaluation} we evaluate the effectiveness and various other metrics of DAMIA, and we also explore factors of the source domain that affect the accuracy of DAMIA. 
In \Cref{sec:related_works}, we present related works and conclude the paper in \Cref{sec:conclusion}. 


\section{Background} 
\label{sec:background}

This section introduces background knowledge including the membership inference attack and  domain adaptation. 


\subsection{Membership Inference Attacks} 
\label{sub:membership_inference_attacks}

Membership inference attack is a type of attack against deep learning models, which can be deployed to determine whether a sample is from the training set of a victim model. 
The basic idea of the attack is that the information exposed by the model contains the abundant information of the training data, based on which an attacker may perform membership inferences. Theoretically, all characteristics of the victim model such as activation values, affine outputs, gradients or even the model's transparency report, can be utilized by attackers to deploy the attack \cite{nasr2018comprehensive,shokri2019privacy,salem2018ml}. Given that most of the above characteristics are not publicly accessible, attackers may solely rely on the outputs of the model to deploy the attack in practice. 

Mathematically, a victim DL model $M$ trained on a dataset $D$, which handles a classification task, categorizing all its inputs into $n$ categories. Any sample $s_i$ fed into $M$ will result in an output $o_i = (p_1,p_2,...,p_j,...,p_n)$, where $p_j$ is the confidence score, indicating the probability of being a sample of the category $j$. To deploy the membership inference attack, an attacker may train a binary classifier $M_{adv}$ to determine whether a sample $s_i$ is from the training data $D$.    
$$
    member\_or\_not = M_{adv}(o_i)
$$
A simple but efficient way to initiate a membership inference attack is to build a linear binary classifier $M_{adv}^{lin.}$ using a confidence threshold $P_{thresh}$ as a decision boundary, which is selected based on extensive experiments. In particular, one may enumerate all possible confidence scores to find the $P_{thresh}$, which can achieve the maximum accuracy of membership inference \cite{yeom2018privacy}. With $P_{thresh}$, if the confidence score of a sample for a specific category $k$ ($1 \leq k \leq n$) is higher than $P_{thresh}$, $M_{adv}^{lin.}$ will determine the sample as a member of the training set. Formally, the procedure can be defined as follows:
$$
    M_{adv}^{lin.}=\left\{
    \begin{aligned}
    1   & &  p_k \geq P_{thresd}  \\
    0   & &  p_k < P_{thresd} 
    \end{aligned}
    \right.
$$
Noted that $1$ stands for member and $0$ stands for non-member.

\killme{
\begin{figure}
    \label{img:membership_inferece}
    \centering
    \includegraphics[width=0.3\textwidth]{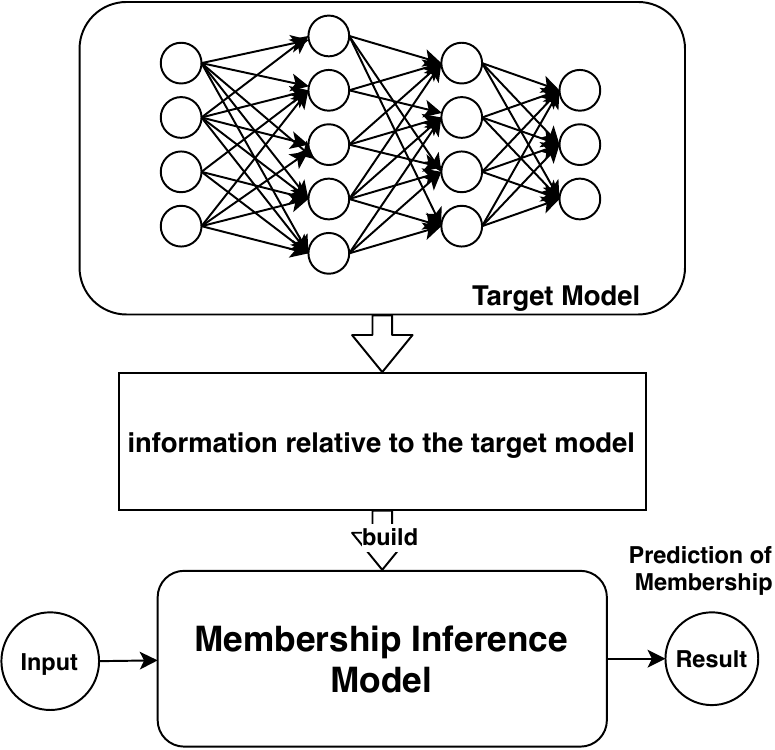}
    \caption{High level idea of membership inference. Information relative to the target model can be the direct output of the model, the gradients during traning or the transparency reports of the target model.}
\end{figure}
}



Compared with other attacks such as model inversion \cite{fredrikson2015model,yang2019neural}, the impact of membership inference attack is relatively minor. 
However, membership inference is easier to deploy and requires less information on the victim model compared with other attacks. Therefore, when an attacker attempts to compromise a model so as to derive the sensitive information, the membership inference attack is served as a ``metric'', probing whether the model is potentially vulnerable. If the membership inference attack is possible, other attacks with higher severity can then be launched.

\subsection{Domain Adaptation}
\label{sub:domain_adaptation}

Domain Adaptation (DA) is a branch of transfer learning \cite{pan2009survey}, aiming to address the issue of insufficient labeled training data. It utilizes the knowledge of one or more relevant source domains to conduct new tasks in a target domain. Mathematically, we denote a domain as $ \mathcal{D} = \{ \mathcal{X} , P(X)\} $, in which $ \mathcal{X} $ represents the feature space and $ P(X) $ represents the margin probability distribution, $ X = \{ x_1, x_2, ... , x_n \} \in \mathcal{X} $. A task on a specific domain is denoted as $ \mathcal{T} = \{ \mathcal{Y}, f(x) \}$, where $ \mathcal{Y} $ is the label space and $ f(x) $ is the target prediction function. Therefore, a source domain can be represented as  $ \mathcal{D}_s = \{ \mathcal{X}_s , P(X)_s\} $, while a target domain  can be represented as $ \mathcal{D}_t = \{ \mathcal{X}_t , P(X)_t \} $. Correspondingly,  $ \mathcal{T}_s = \{ \mathcal{Y}_s, f(x)_s \} $ and $ \mathcal{T}_t = \{ \mathcal{Y}_t, f(x)_t \}$ are two tasks. The goal of DA is to leverage the latent knowledge from $\mathcal{D}_s$ and $ \mathcal{T}_s $ to improve the performance of $ f(x)_t $ in $ \mathcal{T}_t $, where $ \mathcal{D}_s \neq \mathcal{D}_t $. Please noted that in domain adaptation, $ \mathcal{T}_s = \mathcal{T}_t$. 
Basically, the approach achieves the knowledge transferring by driving the model to learn the shared representation of the source domain and target domain. Various approaches of DA have been introduced, which can be grouped into three categories \cite{wang2018deep}, including discrepancy-based approaches, adversarial-based approaches and reconstruction-based approaches. We now elaborate on each category in detail:  


\killme{
\begin{figure}
    \centering
    \label{fig:idea_of_domain_adaptaion}
    \includegraphics[width=0.3\textwidth]{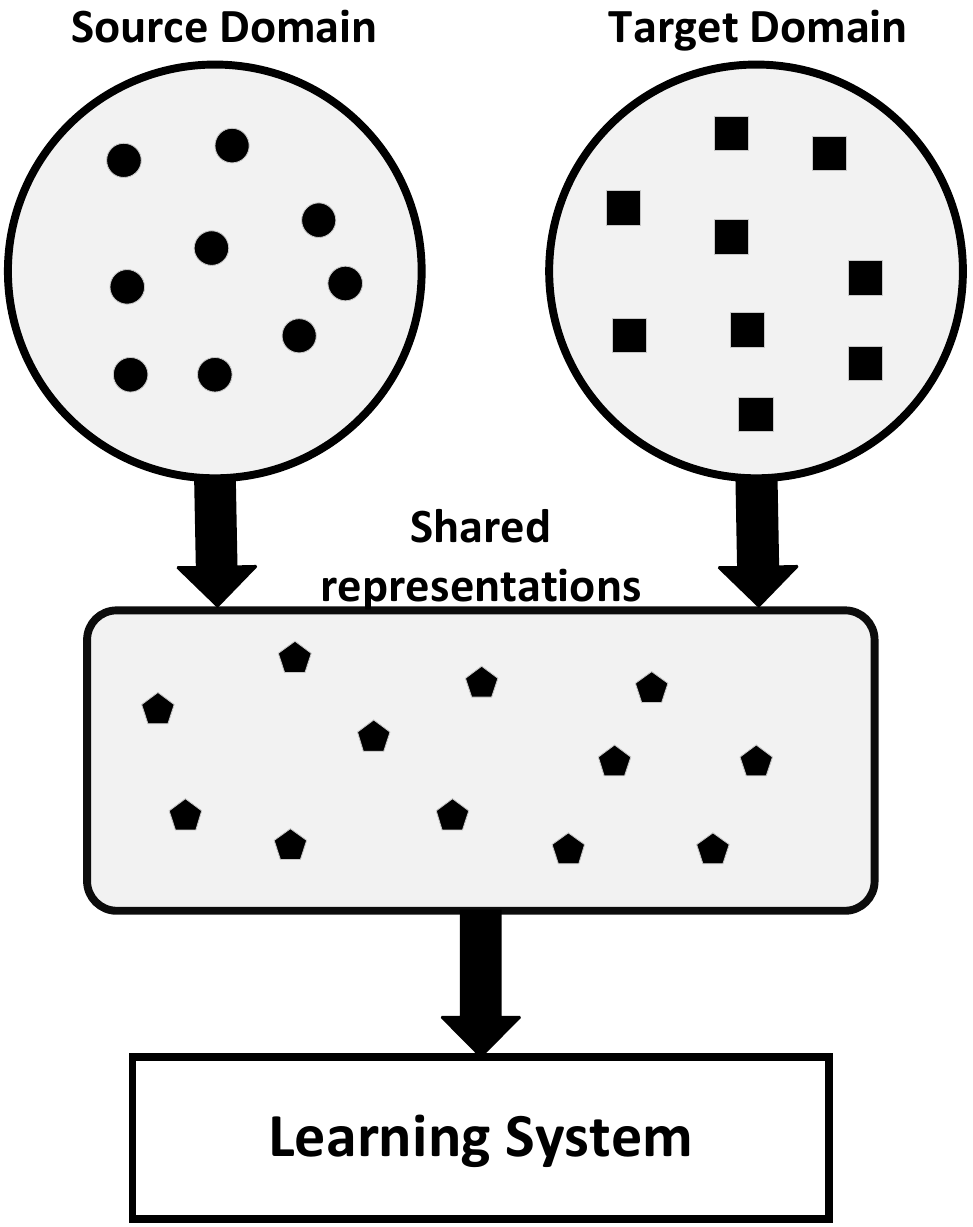}
    \caption{Diagram of Domain Adaptation. Domain adaptation methods find the shared representation of the source and target domain and use this representation to train a learning system} 
\end{figure}
}

\textbf{Discrepancy-based Approaches.}  Discrepancy-based approaches assume that a shared representation of the source and target domain can be obtained by fine-tuning the DL model with labeled or unlabeled data. The approaches can be implemented in multiple ways, and as one of the popular criterion, statistic criterion using some mechanisms to align the statistical distribution shift between the source and target domains. 
For example, \citet{tzeng2014deep} proposed deep domain confusion (DDC) by introducing Maximum Mean Discrepancy (MMD) to the loss function. Specifically, they add MMD into the loss function and regrade it as a metric to quantitatively measure the distance between domains. That is, a smaller enough MMD will reflect that the shared representation has been obtained by the model, and the goal during the training process is to minimize MMD. We give the definition of MMD as follows, please noted that $\phi(\cdot)$ is a kernel function that maps inputs to a space of higher dimensionality:    

\begin{footnotesize}
$$
    D_{\mathcal{M M D}}\left(X_{S}, X_{T}\right)=\left\|\frac{1}{\left|X_{S}\right|} \sum_{x_{s} \in X_{S}} \phi\left(x_{s}\right)-\frac{1}{\left|X_{T}\right|} \sum_{x_{t} \in X_{T}} \phi\left(x_{t}\right)\right\|
$$
\end{footnotesize}
Correspondingly, the loss function can be represented as follows: 
 
$$
    Loss=Loss_{C}\left(X_{L}, y\right)+\lambda D_{\mathcal{M M D}}^{2}\left(X_{S}, X_{T}\right)
$$

\textbf{Adversarial-based Approaches.} 
Adversarial-based approaches construct the shared representation in an adversarial way \cite{tzeng2017adversarial,ajakan2014domain,ganin2017domain}. Specifically, Generative Adversarial Learning (GAN) pit two networks against each other -- known as the discriminator and the generator. Originally, the generator is trained to produce samples that may confuse the discriminator, so that the discriminator may fail to distinguish a generated sample and a sample from the genuine dataset. In Adversarial-based approaches, the principle is adopted to ensure that the discriminator cannot determine whether a sample is generated from a generator, or originally from the source/target domain, indicating the generator is capable of generating the shared representation.

\textbf{Reconstruction-based Approaches.} 
Reconstruction-based approaches construct the shared representation through a reconstruction process. That is, providing that a representation can be reconstructed into samples in both the source and target domains, the representation is considered as the one shared of the source and target domains. Similar to the generator in adversarial-based approaches, a model that can produce such a representation is the one of interest. For example, \citet{ghifary2016deep} uses an autoencoder \cite{vincent2010stacked} for domain adaptation, in which the encoder is for shared representation learning and a decoder is used to reconstruct the representation.

\section{DAMIA: Domain Adaptation against Membership Inference Attacks} 
\label{sec:leveraging_domain_adaptation_as_a_defense} 
\vspace{-1mm}

In this section, we first define our threat model. Afterwards, we shed light on the idea of DAMIA, in which we leverage DA as a defense against membership inference attacks.

\subsection{Threat Model}
\label{subsec:threat_model}
\vspace{-1mm}

DAMIA works under the scenario of MLaaS (Machine-Learning-as-a-Service) where a DL model is served to the public, and only the prediction APIs are exposed to users for feeding inputs. In such a scenario, users can collect the outputs (i.e., the prediction vectors) when the model finishes the processing, and the goal of DAMIA, as the name implies, is to defeat the membership inferring attacks. 

Specifically, we make the following assumptions:

\begin{enumerate}
\item We assume that the attacker has only black-box access to the victim model. That is, the only way an attacker can interact with the victim model is to invoke the prediction APIs and collect outputs from the victim model. 
\item We assume that the attacker knows the distribution that the training data of the victim model drawn from, meaning that an attacker may obtain all the possible values in the training data that the model trained on. However, given specific data, the attacker does not know whether they are in the training dataset. Note that this assumption is also made in most of the membership inference threat model \cite{shokri2017membership, yeom2018privacy, salem2018ml}.

\item We assume that the attacker is not aware of the implementation of the target model, including the architecture, the training algorithm and the hyperparameters such learning rates. This is reasonable because other than the model, the underlying architecture as well as the hyperparameters are usually not public in practical. 
\item We further assume that the attacker has the ability to access the target domain as well as the source domain. 
\end{enumerate}

\subsection{Insight and Design}

Membership inference attacks are feasible due to the fact that the model may achieve a better performance when the input is from the training dataset, which is considered as sensitive. Therefore, a straightforward solution is to obfuscate the sensitive dataset, so that membership inference attacks against the sensitive dataset are not possible. However, a model is trained on an obfuscated dataset, may not achieve the same goal in a specific task as the sensitive dataset does. For example, categorizing pictures of cats and dogs clipped from animation movies may not be performed if the pictures of cats and dogs aforementioned were blurred during the training process. Therefore, exploring a solution that may bridge the gap between the original sensitive dataset and an obfuscated dataset in terms of balancing usability and security is a vital challenge.

\begin{figure}
   \centering
    \includegraphics[width=0.48\textwidth]{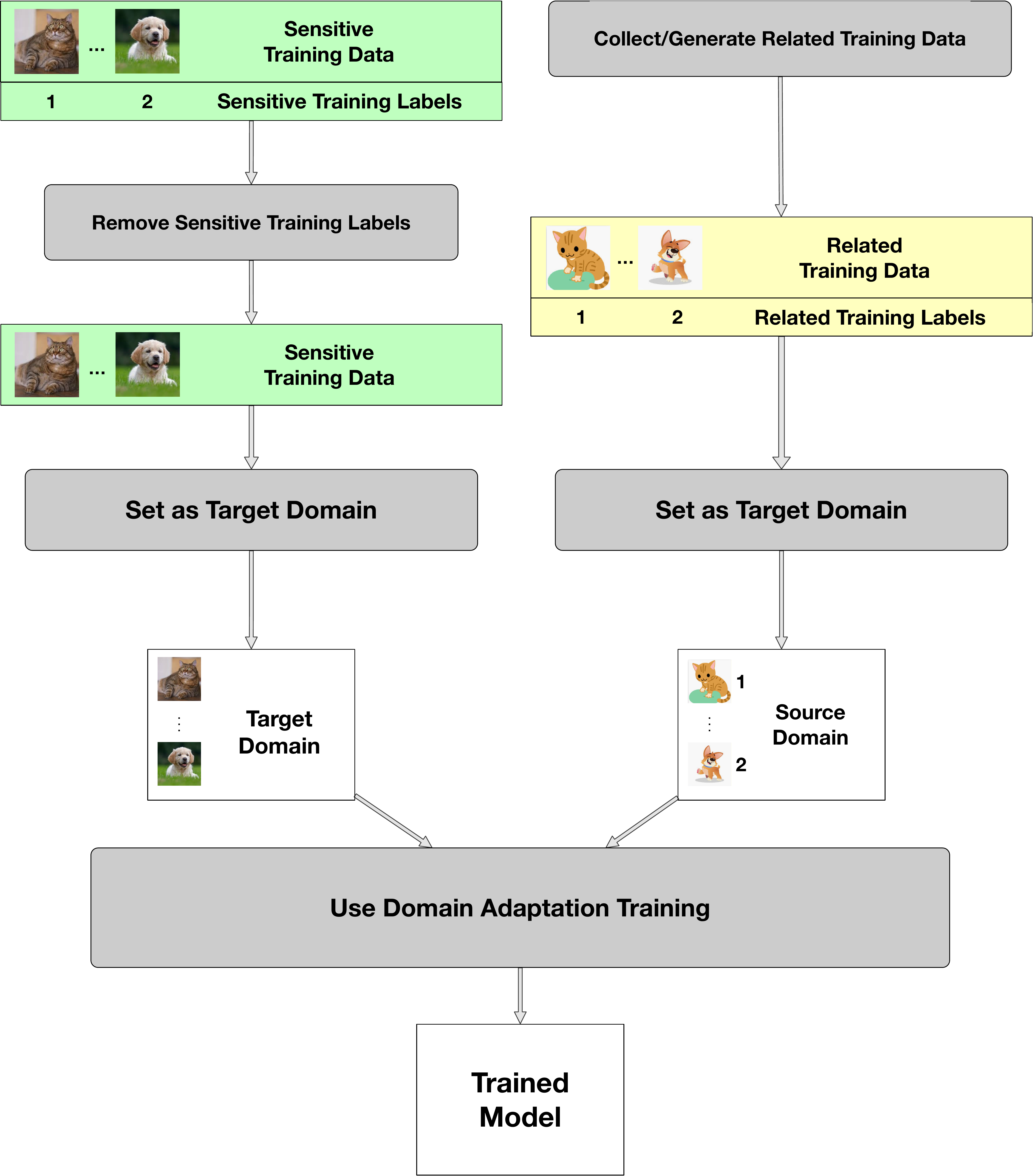}
    \caption{Workflow of DAMIA.}
     \label{fig:overview_of_damia}
\end{figure}

In this paper, we propose DAMIA, i.e., Domain Adaptation against Membership Inference Attacks. As the name implies, DAMIA adopts the Domain Adaptation as the core engine of our solution. Our intuition is that Domain Adaptation can generate a shared representation (i.e., a model) of a source domain and a target domain, where the sensitive dataset belongs to the target domain, and the other dataset from source domain is provided to obfuscate the sensitive data set. As a consequence, the shared representation should have good support in defending membership inference attack against a sensitive dataset, since the sensitive dataset is obfuscated during the training process. Meanwhile, the extracted shared representation can also be used to solve tasks that require the use of the sensitive dataset, which has been widely discussed in \cite{tzeng2017adversarial,ajakan2014domain,ganin2017domain}.

\autoref{fig:overview_of_damia} illustrates the workflow, which includes the flowing steps: 
\begin{enumerate}
\item Domain adaptation requires one dataset from the source domain and one dataset from the target domain. Initially, only the sensitive data is given, and the domain that the sensitive dataset belongs to is now referred to the target domain. For example, the sensitive dataset in our context can be an image-set which contains pictures of cat and dogs clipped from animation movies.
\item We find a dataset other than the sensitive dataset in a different but related domain. For example, we can find an image-set collected from Instagram, containing pictures of cats and dogs from the real world. 
\item we adopt domain adaptation training to train a model. In our exemplary example, domain adaptation will train a shared representation based on the two image-sets containing images of cats and dogs, i.e., image-set collected from animation movies and image-set collected from Instagram. Please note that the labels of the first image-set are removed before the training to cohere with the training process of domain adaptation. Consequently, the shared representation will not raise a violation of confidential of the image-set collected from animation movies, which is considered sensitive. 
\end{enumerate}
 

\vspace{-2mm}
\section{ Evaluation }
\label{sec:evaluation}
\vspace{-1mm}

We now explore analytically the performance of DAMIA. As stated above, the intuition is that DA can obfuscate sensitive datasets with a different but related dataset, generating a shared representation that shares underlying common features with the sensitive dataset, thwarting the membership inference attack. Obviously, there is a trade-off between security and usability, due to the fact that an obfuscated dataset may not achieve the same performance as the original dataset does. Given that, we attempt to answer the following three questions. 
\vspace{2mm}
\begin{tcolorbox}[label=question]
\textbf{Q1}:\textit{ Is DAMIA effective in countering membership inference attack?  
 }
 
 \textbf{Q2}:\textit{ Do all types of domain adaptation techniques have the same effects in countering membership inference attack?
 }
  \textbf{Q3}:\textit{ What are the factors that may impact the performance of DAMIA? And how could the DAMIA achieve the best performance by manipulating those factors? 
 }
\end{tcolorbox}

 The rest of this section is organized as follows: we first present our experiment setup. Afterwards, we use three standalone subsections to evaluate different aspects of DAMIA, answering the aforementioned three questions correspondingly. 


\subsection{ Experiment Setup }
\label{sec:setup }

All our experiments are conducted on a Ubuntu 16.04 server with a Intel(R) Xeon(R) CPU E5-2640 CPU, 4 GTX 1080 GPUs and memory with a size of 130 GB. We use PyTorch \footnote{https://github.com/pytorch/pytorch} to build our deep learning models for evaluation. The datasets involved in our experiments including ($i$) \texttt{MINST} ($ii$) \texttt{SVHN} and ($iii$) \texttt{Office-31}. 

\begin{figure}  
    \centering
    \includegraphics[width=0.45\textwidth]{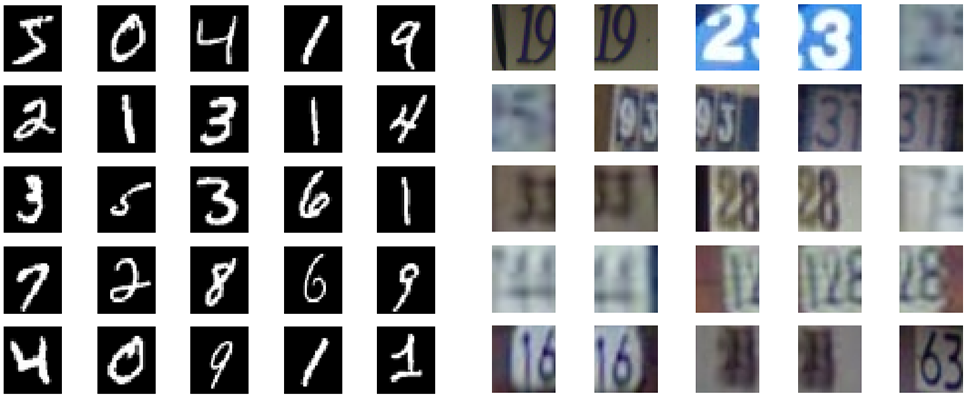}
    \caption{Representative samples from \texttt{MNIST} (left) and \texttt{SVHN} (right).}
    \label{fig:mnist_SVHN}
\end{figure}

\begin{figure} 
    \centering
    \includegraphics[width=0.45\textwidth]{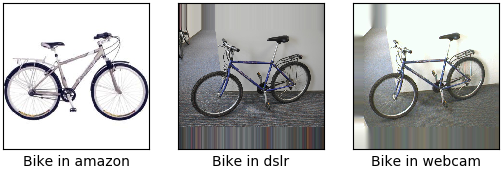}
    \caption{Images of bikes from 3 different domains in \texttt{Office-31}.}
    \label{fig:office31}
\end{figure}

Specifically, \texttt{MINST} is an image dataset containing handwritten digits with an image size of $ 28 \times 28 $. All the digits are in the range of 0 to 9 and all the images are gray-scale. \texttt{MNIST} contains 70,000 images totally, in which 60,000 images are used as the training set and the rest are served as the non-training set (i.e. test set).
Similar to \texttt{MNIST}, \texttt{SVHN} is also an image dataset of digits of 0 to 9 with an image size of $ 32 \times 32 $. The training dataset contains 73257 images, the non-training dataset is with a size of 26032, and the extra training dataset is with a size of 531131.  Some representative samples from the two datasets are shown in \autoref{fig:mnist_SVHN}. \texttt{Office-31} \footnote{https://people.eecs.berkeley.edu/~jhoffman/domainadapt/} is a dataset about objects commonly appeared in an office setting. The dataset totally consists of 4110 images from three domains --- \texttt{Amazon}, \texttt{DSLR} and \texttt{Webcam}, each domain contains 31 categories respectively, which is designed for domain adaptation. Images in domain \texttt{Amazon} are collected directly from amazon.com, which are prepossessed so that there is only a target object in a blank white background. Images in domains \texttt{Webcam} and \texttt{DSLR} are shot by web cameras and DSLR (digital single-lens reflex) cameras in a real-world office. The two domains are similar to each other, and the most difference is the object's pose and the lighting condition. The details of \texttt{Office-31} is are shown in \autoref{tab:details_of_office_31} and the difference among the three domains are shown in \autoref{fig:office31}. Please note that all the datasets are utilized as either source domain or target domain, and are split into training sets and non-training sets.

\begin{table} 
    \centering
    \caption{Details of \texttt{Office-31}}
    \label{tab:details_of_office_31}
    \resizebox{0.48\textwidth}{!}{%
    \begin{tabular}{cccc}
    \toprule
    \textbf{Domains} & \textbf{Total Amount} & \textbf{Training Set (80\%)} & \textbf{Non-training Set (20\%)} \\
    \midrule
    \textbf{Amazon} & 2817  & 2253                & 564             \\
    \rowcolor{table_gray}
    \textbf{DSLR  } & 498   & 398                 & 100             \\
    \textbf{Webcam} & 795   & 636                 & 159             \\
    \bottomrule
    \end{tabular}
    }
\end{table}

\subsection{Effectiveness }
\label{sub:effectiveness}

To \textbf{Q1}, we define the effectiveness of DAMIA as the capabilities in defending membership inference attacks. We use two models with the same architecture, i.e., AlexNet \cite{krizhevsky2012imagenet}, and adopt a few metrics to demonstrate it. The two models are: ($i$) a model trained on dataset \texttt{Webcam} from \texttt{Office-31}; ($ii$) a model trained by DAMIA, in which the target domain and source domain may vary according to the specific context.  
During the training process, the discrepancy-based domain adaptation technique, Deep Domain Confusion \cite{tzeng2014deep} (DDC), is adopted for training DAMIA.
The membership inference attack used in our experiment is a threshold-based membership inference attack introduced in \cite{yeom2018privacy,song2019privacy}. We will not provide the details of the attack due to the page limit.

The origination of this part is as follows: initially, we evaluate the effectiveness of DAMIA by using two legacy metrics, which will reflect the advantage of an attacker. In particular, the metrics introduced in our paper are also widely adopted in other efforts \cite{nasr2018machine,yeom2018privacy}, and these metrics are generalization error, prediction distributions. 
Further, we also introduce two novel metrics, which may also reflect the capabilities of a model in defending the membership inference attacks.  These metrics are the intermediate representation and the advantage of membership inference attacks.
Please refer to the corresponding paragraph for the definition of each metric.

\subsubsection{Legacy Metrics of Effectiveness} 

\begin{figure} 
    \centering
    \includegraphics[width=0.45\textwidth]{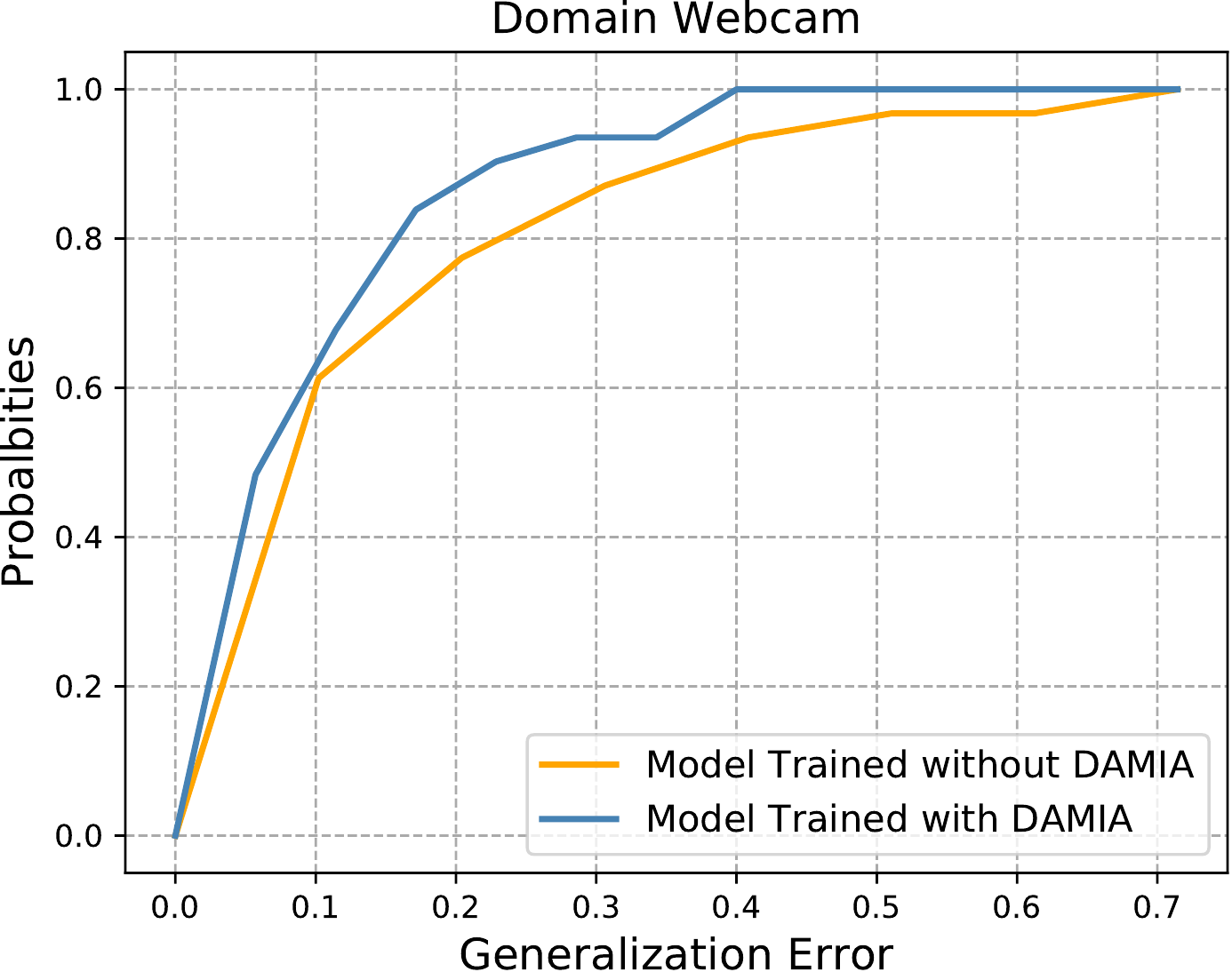}
    \caption{Empirical Cumulative Distribution Function (CDF) of the generalization error of models across different categories in \texttt{Webcam}.   \texttt{Amazon} is used for obfuscation.}
    \label{fig:alys_gene_err1}
\end{figure}

\begin{figure} 
    \centering
    \includegraphics[width=0.45\textwidth]{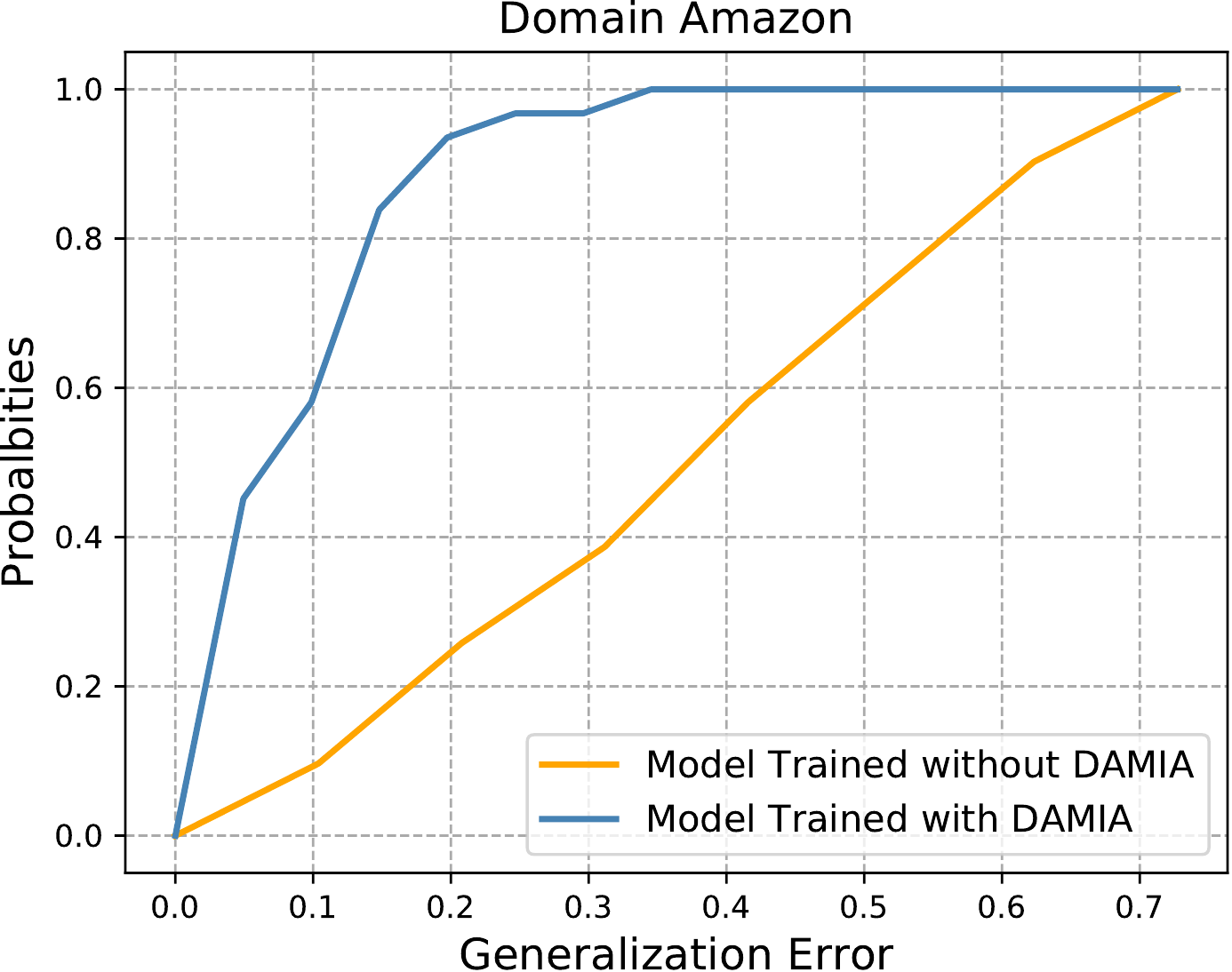}
    \caption{Empirical Cumulative Distribution Function (CDF) of the generalization error of models across different categories in \texttt{Amazon}.  \texttt{Webcam} is used for obfuscation.  }
    \label{fig:alys_gene_err2}
\end{figure}

\textbf{Generalization Error}:  Generalization error is a metric to evaluate the performance of a model when a non-training dataset is involved. If the generalization error is low, a model can achieve a good performance, which may be close to the performance when the model tests on the training dataset. Therefore, an attacker may want to maximize the generalization error. For each dataset, we trained two models including one with DAMIA enabled, while the other one is idle. For the one with DAMIA enabled, we first assume the attacker may want to deploy attacks against \texttt{Webcam}, so that a dataset from \texttt{Amazon} is used in our experiment to obfuscate \texttt{Webcam}. We then assume the attacker may want to attack against \texttt{Amazon}, and similarly, the \texttt{Webcam} is then used to obfuscate \texttt{Amazon}. 
\autoref{fig:alys_gene_err1} and \autoref{fig:alys_gene_err2} show the generalization error for \texttt{Webcam} and \texttt{Amazon} respectively. It can be observed that whatever the dataset is,  a model with DAMIA enabled has a lower generalization error when compared with the one without DAMIA, indicating the DAMIA is effective in defending against the membership inferences attacks.

\begin{figure} 
    \centering
    \includegraphics[width=0.48\textwidth]{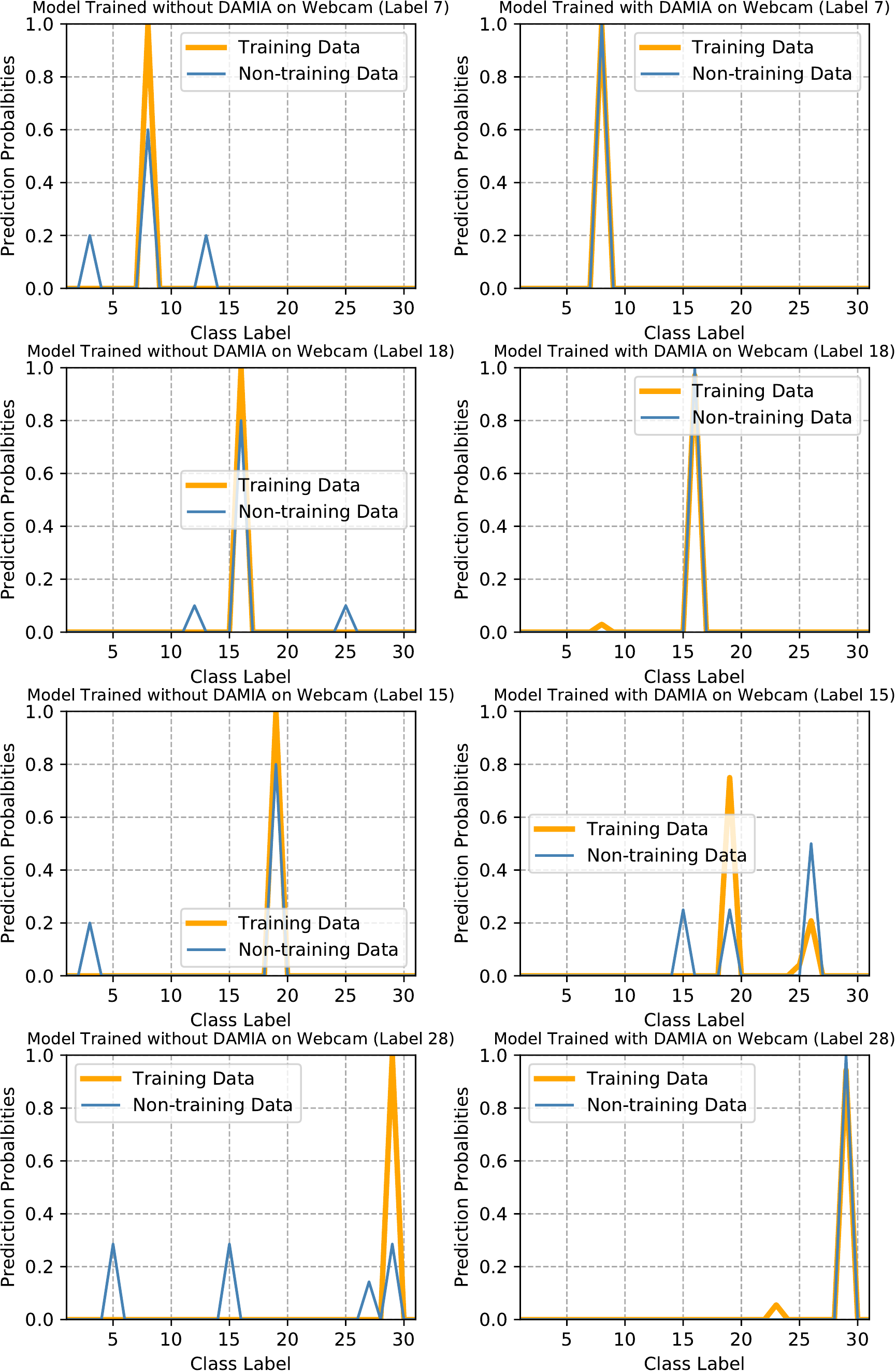}
    \caption{Prediction distribution comparison.}
    \label{fig:alys_ditribution_cls}
\end{figure}

\textbf{Prediction Distributions}:  Prediction distribution is referred to as the distribution of the probability of being a sample from each category \cite{nasr2018machine} \killme{(i.e., training dataset and non-training dataset)}. Basically, if the prediction distribution of a training dataset and that of the non-training dataset are close to each other, one cannot tell difference. In other words, the membership inference attack may fail in this case. The experiment procedures and dataset used are similar to the first one, and we will not go to details. 
Recall that each dataset in \texttt{Office-31} contains different categories of images. We select and plot the prediction distributions of a few categories, including desk chair (category label ``7''), mobile phone (category label ``15''), ring binder (category label ``25'') and speaker (category label ``28'').  As demonstrated in \autoref{fig:alys_ditribution_cls}, the figures on the left side show categories of the model with DAMIA enabled, while the ones on the left side that does not. It can be observed that the prediction distributions of the model with DAMIA are more approximate when compared with the one without. This will bring more challenges for an attacker whose goal is to deploy the membership inference attacks.


\subsubsection{New Metrics of Effectiveness}

\begin{figure*} 
    \centering
    \includegraphics[width=0.8\textwidth]{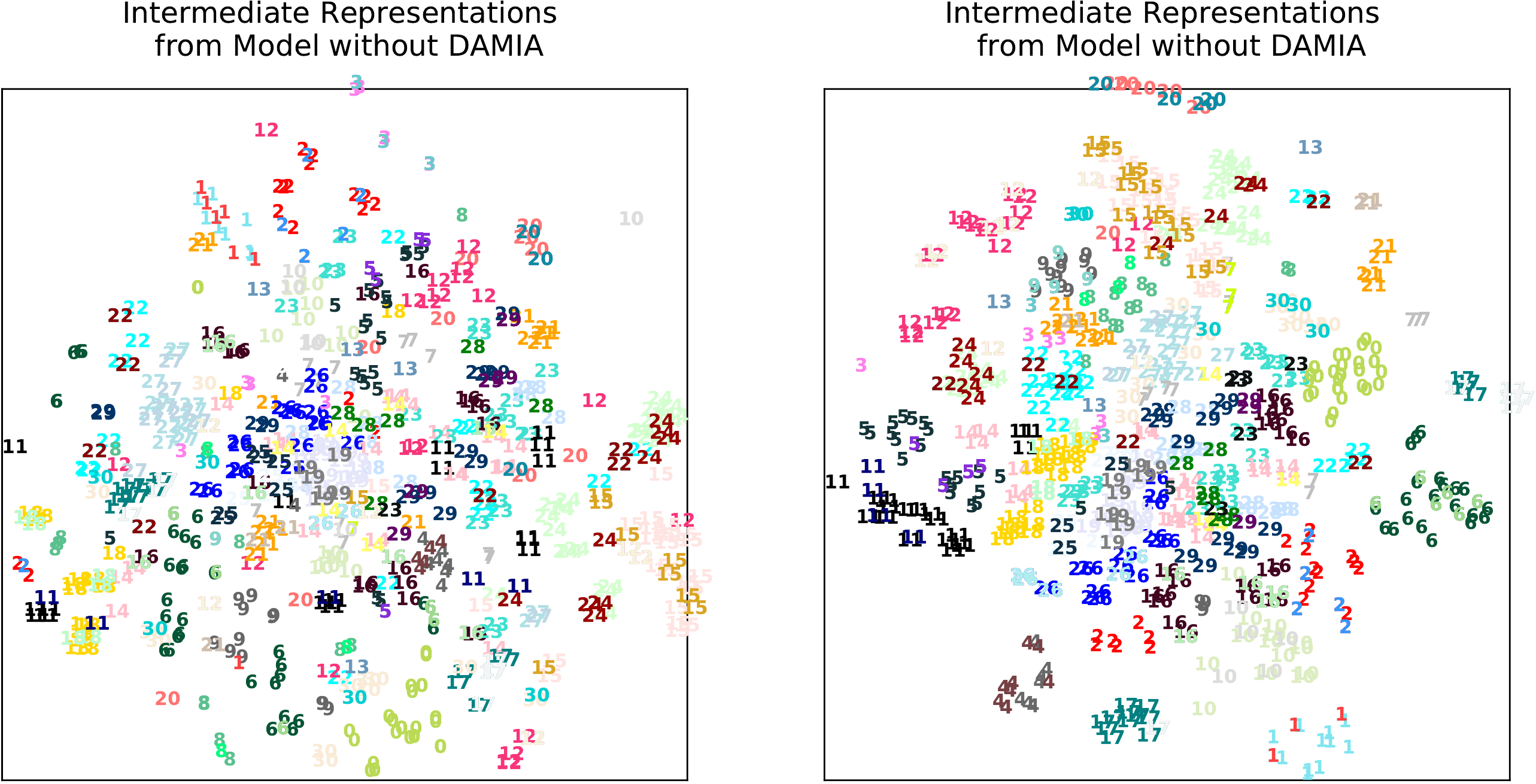}
    \caption{Intermediate representations of \texttt{Webcam} processed by the models without (left) or with (right) DAMIA enabled.}
    \label{fig:alys_features}
\end{figure*}

\textbf{Intermediate Representation}: Intermediate representation reveals the way of a DL model processes its training data. Basically, a DL model will go through a feature extraction process and a classification process throughout the entire training, and the outputs of the feature extraction process are referred to as intermediate representations, which may affect the accuracy of the classification process. Given that an attacker leverage the outputs of the classification process to derive membership inference attack, intuitively, the intermediate representation may also have an association with the success rate of deploying the membership inference attack. We conduct a similar experiment to confirm our intuition, and the intermediate representations are from the fifth convolution layer (the training process goes through multiple training layers \cite{lecun2015deep}), with their dimensionalities (i.e., the number of attributes that data have) reduced to two \cite{maaten2008visualizing}.  Intermediate representations are shown in \autoref{fig:alys_features}. Samples from the same category are with the same color and the same category label, and we carefully assign all the colors to avoid the conflicts. It can be observed that, with DAMIA enabled, data of each category forms a more compact cluster, and the intermediate representations of the non-training samples are much close to the training samples.  
The result indicates that the way that the model trained by DAMIA does process the training data and non-training in a similar way, creating barriers for distinguishing training data from non-training data. In other words,   membership inference attacks are more likely to fail.

\textbf{Advantage of an Adversary}: We now introduce a new metric to measure the advantage of an adversary who wants to deploy a membership inference attack against a model.  With insufficient advantage, the accuracy of membership inference is merely close to a random guess. 
Therefore, the membership inference advantage of an adversary $adv._{mi}$ can be defined as:   
$$
    adv_{mi} = = \frac{|P_{inference} - P_{random}|}{P_{random}}
$$
In particular, $P_{inference}$ represents the accuracy of membership inference for a specific adversary, while $P_{random}$ is the probability of a random guess, which is 0.5, in that for a specific sample, an adversary without any foreknowledge may randomly output yes or no to guess whether the sample is from the training dataset. For example, when $adv_{mi} $ closes to zero, the membership inference attack merely a random guess. 
We believe the metrics can better reflect the capability when compared with the legacy ones, in that it gives a result indicating whether the attack works in a straightforward manner. 
Further, to demonstrate the feasibility, we adopt the metric $adv_{mi}$  to measure the advantages in practice and illustrate the results in \autoref{tab:adv_w_wout_DAMIA}. As expected, with DAMIA, the advantages of attackers are significantly hindered. Please note that the training process, as well as other related information of our testing models, have been introduced before, and please refer to the previous subsection for details.  
 
\begin{table}[H]
    \caption{
    Advantages comparison on \texttt{Amazon} and \texttt{Webcam} with/without DAMIA enabled. Please note that ``Acc.'' is referred to ``Accuracy'', and ``MIA'' is referred to membership inference attack. }
    \label{tab:adv_w_wout_DAMIA}
    \setlength{\tabcolsep}{1mm}{
    \begin{tabular}{ccccc}\hline
    \toprule
    \multirow{2}{*}{\textbf{Sensitive data}} & \multicolumn{2}{c}{\textbf{Amazon}}          & \multicolumn{2}{c}{\textbf{Webcam}} \\ 
    \cmidrule{2-5}
                   & \textbf{w/o DAMIA}     & \textbf{DAMIA}                   & \textbf{w/o DAMIA}    & \textbf{DAMIA}          \\ 
    \midrule
    \textbf{MIA Acc.}        & 0.77324    & 0.514514926             & 0.68003    & 0.534591195    \\
    \rowcolor{table_gray}
    \textbf{$adv._{mi}$}    & 0.54648    & \textbf{0.029029852}    & 0.36006    & \textbf{0.06918239}     \\ 
    \bottomrule
    \end{tabular}
    }
\end{table}

\subsection{Performance of different domain adaptation approaches}
\label{sub:feasibility_of_different_domain_adaptation_approaches}

To \textbf{Q2}, we explore if the use of a different domain adaptation technique will affect the performance of DAMIA. To this end, we train DAMIA with all the three approaches introduced in Section \ref{sec:background} respectively, and evaluate the corresponding performance of each model in terms of defending against membership inference attack. As discussed before, using advantage $adv_{mi}$ is a simple but effective way to evaluate if the membership inference attacks are possible. Therefore, we will continue using this metric in the rest of this section.

\textbf{Discrepancy-based Domain Adaptation}: We evaluate the feasibility for DAMIA that is trained via discrepancy-based domain adaptation approaches. Similar to the previous experiment, DDC (deep domain confusion \cite{tzeng2014deep} is adopted. We then deploy membership inference attacks against \texttt{MNIST}, \texttt{SVHN} and the three datasets in \texttt{Office-31}. To eliminate the impact of the architecture of a DL model, we use AlexNet and ResNet-50 \cite{he2016deep} as the backbones of the models respectively and have them trained for 150 epochs to make sure that they are converged. The experiments are shown in \autoref{tab:exp01_mnist_svhn_alexnet} to \autoref{tab:exp01_offcie31_resnet50}. It can be observed DAMIA works against membership inference attacks effectively despite the different architectures.

\begin{table} 
    \centering
    \caption{Performance comparison on \texttt{MNSIT} and \texttt{SVHN} (with AlexNet).  Please note that ``Acc.'' is referred to ``Accuracy'', and ``MIA'' is referred to membership inference attack. The same below.  }
    \label{tab:exp01_mnist_svhn_alexnet}
    \centering
    \resizebox{0.48\textwidth}{!}{%
    \begin{tabular}{cccccccc}
        \toprule
        \tabincell{c}{\textbf{Adapation}\\ \textbf{Direction (DAMIA)}}    & \tabincell{c}{\textbf{Train Acc.}\\ \textbf{on Target}}  &  \tabincell{c}{\textbf{Test Acc.}\\ \textbf{on Target}}  &  \tabincell{c}{\textbf{MIA Acc.}\\ \textbf{on Target}} & \tabincell{c}{\textbf{$adv_{mi}$}\\ \textbf{on Target}} \\  
        \midrule
        \textbf{MNIST} $\rightarrow$ \textbf{SVHN}  & 0.212566717 & 0.224953903  & \textbf{0.503727767} & \textbf{0.007455534} \\ 
        \rowcolor{table_gray}
        \textbf{SVHN} $\rightarrow$ \textbf{MNIST}  & 0.76805     & 0.7702       & \textbf{0.503108333} & \textbf{0.006216666} \\ 
        \bottomrule 
        \end{tabular}%
    }
\end{table}

\begin{table} 
    \centering
    \caption{Performance comparison on \texttt{Office-31} (with AlexNet).  }
    \label{tab:exp01_offcie31_alexnet}
    \centering
    \resizebox{0.48\textwidth}{!}{%
    \begin{tabular}{cccccccc}
    
    \toprule
        \tabincell{c}{\textbf{Adapation}\\ \textbf{Direction (DAMIA)}}    & \tabincell{c}{\textbf{Train Acc.}\\ \textbf{on Target}}  &  \tabincell{c}{\textbf{Test Acc.}\\ \textbf{on Target}}  &  \tabincell{c}{\textbf{MIA Acc.}\\ \textbf{on Target}} & \tabincell{c}{\textbf{$adv_{mi}$}\\ \textbf{on Target}} \\  
        \midrule
    \textbf{Amazon} $\rightarrow$ \textbf{DSLR}   & 0.447236181 & 0.46         & \textbf{0.528492462} & \textbf{0.056984924} \\ 
    \rowcolor{table_gray}
    \textbf{Amazon} $\rightarrow$ \textbf{Webcam} & 0.501572327 & 0.459119497  & \textbf{0.534591195} & \textbf{0.06918239} \\ 
    \textbf{DSLR} $\rightarrow$ \textbf{Amazon}   & 0.407900577 & 0.416666667  & \textbf{0.510169656} & \textbf{0.020339312} \\ 
    \rowcolor{table_gray}
    \textbf{DSLR} $\rightarrow$ \textbf{Webcam}   & 0.938679245 & 0.93081761   & \textbf{0.522012579} & \textbf{0.044025158} \\ 
    \textbf{Webcam} $\rightarrow$ \textbf{Amazon} & 0.375055482 & 0.372340426  & \textbf{0.514514926} & \textbf{0.029029852} \\ 
    \rowcolor{table_gray}
    \textbf{Webcam} $\rightarrow$ \textbf{DSLR}   & 0.929648241 & 0.91         & \textbf{0.544296482} & \textbf{0.088592964} \\ 
    \bottomrule 
        \end{tabular}%
    } 
\end{table}

\begin{table} 
    \centering
    \caption{Performance comparison on \texttt{MNSIT} and \texttt{SVHN} (with ResNet-50).}
    \label{tab:exp01_mnist_svhn_resnet50}
    \centering
    \resizebox{0.48\textwidth}{!}{%
    \begin{tabular}{cccccccc}
        
        \toprule
        \tabincell{c}{\textbf{Adapation}\\ \textbf{Direction (DAMIA)}}    & \tabincell{c}{\textbf{Train Acc.}\\ \textbf{on Target}}  &  \tabincell{c}{\textbf{Test Acc.}\\ \textbf{on Target}}  &  \tabincell{c}{\textbf{MIA Acc.}\\ \textbf{on Target}} & \tabincell{c}{\textbf{$adv_{mi}$}\\ \textbf{on Target}} \\  
        \midrule
        \textbf{MNIST} $\rightarrow$ \textbf{SVHN}  & 0.219173594 & 0.227566073 & \textbf{0.511505038} &\textbf{0.023010076} \\ 
        \rowcolor{table_gray}
        \textbf{SVHN} $\rightarrow$ \textbf{MNIST}  & 0.778966667 & 0.7844      & \textbf{0.500133333} &\textbf{0.000266666} \\ 
        \bottomrule 
        \end{tabular}%
    }
\end{table}

\begin{table} 
    \centering
    \caption{Performance comparison on \texttt{Office-31} (with ResNet-50).}
    \label{tab:exp01_offcie31_resnet50}
    \centering
    \resizebox{0.48\textwidth}{!}{%
    \begin{tabular}{cccccccc}
    
    \toprule
        \tabincell{c}{\textbf{Adapation}\\ \textbf{Direction (DAMIA)}}    & \tabincell{c}{\textbf{Train Acc.}\\ \textbf{on Target}}  &  \tabincell{c}{\textbf{Test Acc.}\\ \textbf{on Target}}  &  \tabincell{c}{\textbf{MIA Acc.}\\ \textbf{on Target}} & \tabincell{c}{\textbf{$adv_{mi}$}\\ \textbf{on Target}} \\  
        \midrule
    \textbf{Amazon} $\rightarrow$ \textbf{DSLR}   & 0.733668342 & 0.68         & \textbf{0.542964824} & \textbf{0.085929648} \\ 
    \rowcolor{table_gray}
    \textbf{Amazon} $\rightarrow$ \textbf{Webcam} & 0.721698113 & 0.761006289  & \textbf{0.517295597} & \textbf{0.034591194} \\ 
    \textbf{DSLR} $\rightarrow$ \textbf{Amazon}   & 0.625832224 & 0.675531915  & \textbf{0.507515983} & \textbf{0.015031966} \\ 
    \rowcolor{table_gray}
    \textbf{DSLR} $\rightarrow$ \textbf{Webcam}   & 0.962264151 & 0.981132075  & \textbf{0.504716981} & \textbf{0.009433962} \\ 
    \textbf{Webcam} $\rightarrow$ \textbf{Amazon} & 0.628051487 & 0.663120567  & \textbf{0.501553484} & \textbf{0.003106968} \\ 
    \rowcolor{table_gray}
    \textbf{Webcam} $\rightarrow$ \textbf{DSLR}   & 0.992462312 & 0.99         & \textbf{0.551909548} & \textbf{0.103819096} \\ 
    \bottomrule 
        \end{tabular}%
    }
\end{table}

\textbf{Adversarial-based Domain Adaptation}:  
We evaluate the feasibility for DAMIA that is trained via adversarial-based domain adaptation approaches. Similar to the first experiment, we deploy membership inference attacks against \texttt{MNIST}, \texttt{SVHN} and the three datasets in \texttt{Office-31}. AlexNet and ResNet-50 \cite{he2016deep} are severed as the backbones of the models respectively. The adversarial-based domain adaptation approach used in the experiment is ADDA (i.e., Adversarial Discriminative Domain Adaptation) \cite{tzeng2017adversarial}.  It can be observed that DAMIA works against membership inference attacks effectively in \autoref{tab:exp02_mnist_svhn_alexnet_adda}.
Unfortunately,  the performances are not as good as we expected. For example, when \texttt{Amazon} is adopted to obfuscate \texttt{Webcam},  the test accuracy of \texttt{Webcam} is down to 1.2579\%. 
Likewise, when \texttt{DSLR} is adopted to obfuscate \texttt{Webcam}, the test accuracy is only 6.918239\%, even when ResNet-50 is used as the backbone.  ResNet-50 has more layers when compared with AlexNet, and theoretically, it should have a better capacity in learning representations \cite{sun2016deep}. This may be attributed to the fact that when GAN is adopted, the model is often difficult to reach convergence \cite{barnett2018convergence}, placing barriers for the model to learn the shared representation. 
We also find that the procedures of training DAMIA via adversarial-based domain adaptation approaches are sophisticated, in that there are four models are involved. Given its unsatisfactory performance, we argue that it is not recommended to train DAMIA via that type of approach.

\begin{table} 
    \centering
    \caption{Performance comparison on \texttt{MNSIT} and \texttt{SVHN} (with ADDA).}
    \label{tab:exp02_mnist_svhn_alexnet_adda}
    \centering
    \resizebox{0.48\textwidth}{!}{%
    \begin{tabular}{cccccccc}
        
        \toprule
        \tabincell{c}{\textbf{Adapation}\\ \textbf{Direction (DAMIA)}}    & \tabincell{c}{\textbf{Train Acc.}\\ \textbf{on Target}}  &  \tabincell{c}{\textbf{Test Acc.}\\ \textbf{on Target}}  &  \tabincell{c}{\textbf{MIA Acc.}\\ \textbf{on Target}} & \tabincell{c}{\textbf{$adv_{mi}$}\\ \textbf{on Target}} \\  
        \midrule
        \textbf{MNIST} $\rightarrow$ \textbf{SVHN}   & 0.137803    & 0.139943  & \textbf{0.5148440} & \textbf{0.029688}   \\ 
        \rowcolor{table_gray}
        \textbf{SVHN} $\rightarrow$ \textbf{MNIST}   & 0.735633    & 0.7426    & \textbf{0.5007183} & \textbf{0.0014366}   \\ 
        \bottomrule 
        \end{tabular}%
    }
\end{table}

\textbf{Reconstruction-based Domain Adaptation}: We further evaluate the feasibility of DAMIA that is trained via reconstruction-based domain adaptation approaches. We deploy membership inference attacks against \texttt{MNIST}, \texttt{SVHN} and the three datasets in \texttt{Office-31}. We choose DRCN (i.e., Deep Reconstruction Classification Networks) \cite{ghifary2016deep} to conduct our experiments. For \texttt{MNIST} and \texttt{SVHN}, we select a vallina CNN with three convolution layers and two fully connected layers as the backbone of the model. Again, we use AlexNet as the backbone for \texttt{Office-31}. Experiment results are shown in \autoref{tab:exp03_mnist_svhn} and \autoref{tab:exp03_office}. Similarly, the advantage for an attacker is approaching zero, indicating that DAMIA work against membership inference attacks effectively.


\begin{table} 
    \centering
    \caption{Performance comparison on \texttt{MNSIT} and \texttt{SVHN} (with DRCN).}
    \label{tab:exp03_mnist_svhn}
    \centering
    \resizebox{0.48\textwidth}{!}{%
    \begin{tabular}{cccccccc}
        
        \toprule
        \tabincell{c}{\textbf{Adapation}\\ \textbf{Direction (DAMIA)}}    & \tabincell{c}{\textbf{Train Acc.}\\ \textbf{on Target}}  &  \tabincell{c}{\textbf{Test Acc.}\\ \textbf{on Target}}  &  \tabincell{c}{\textbf{MIA Acc.}\\ \textbf{on Target}} & \tabincell{c}{\textbf{$adv_{mi}$}\\ \textbf{on Target}} \\  
        \midrule
        \textbf{MNIST} $\rightarrow$ \textbf{SVHN}   & 0.162838     & 0.191726  & \textbf{0.525682} & \textbf{0.051364}    \\ 
        \rowcolor{table_gray}
        \textbf{SVHN} $\rightarrow$ \textbf{MNIST}   & 0.682617     & 0.696000  & \textbf{0.500008} & \textbf{0.000016}    \\ 
        \bottomrule 
        \end{tabular}%
    }
\end{table}

\begin{table} 
    \centering
    \caption{Performance comparison on \texttt{Office-31} (with DRCN).}
    \label{tab:exp03_office}
    \centering
    \resizebox{0.48\textwidth}{!}{%
    \begin{tabular}{cccccccc}
    
    \toprule
        \tabincell{c}{\textbf{Adapation}\\ \textbf{Direction (DAMIA)}}    & \tabincell{c}{\textbf{Train Acc.}\\ \textbf{on Target}}  &  \tabincell{c}{\textbf{Test Acc.}\\ \textbf{on Target}}  &  \tabincell{c}{\textbf{MIA Acc.}\\ \textbf{on Target}} & \tabincell{c}{\textbf{$adv_{mi}$}\\ \textbf{on Target}} \\  
        \midrule
    \textbf{Amazon} $\rightarrow$ \textbf{DSLR}   & 0.412060302 & 0.55        & \textbf{0.568492462} & \textbf{0.136984924} \\ 
    \rowcolor{table_gray}
    \textbf{Amazon} $\rightarrow$ \textbf{Webcam} & 0.375786164 & 0.459119497 & \textbf{0.550314465} & \textbf{0.10062893} \\ 
    \textbf{DSLR} $\rightarrow$ \textbf{Amazon}   & 0.220594763 & 0.242907801 & \textbf{0.526324239} & \textbf{0.052648478} \\ 
    \rowcolor{table_gray}
    \textbf{DSLR} $\rightarrow$ \textbf{Webcam}   & 0.798742138 & 0.836477987 & \textbf{0.58490566 } & \textbf{0.16981132} \\ 
    \textbf{Webcam} $\rightarrow$ \textbf{Amazon} & 0.253883711 & 0.290780142 & \textbf{0.517025762} & \textbf{0.034051524} \\ 
    \rowcolor{table_gray}
    \textbf{Webcam} $\rightarrow$ \textbf{DSLR}   & 0.891959799 & 0.9         & \textbf{0.600452261} & \textbf{0.200904522} \\ 
    \bottomrule 
        \end{tabular}%
    }
\end{table}

\subsection{Factors that Affects the Usability}
\label{sub:factors_of_the_source_domain}

In the previous section, our experiment shows that DAMIA work against membership inference attacks effectively despite the approaches. However, some of the approaches may not have a good test accuracy, which hurts the usability of the original dataset. Therefore, in this section, we want to explore the factors that may affect the usability of DAMIA, which is the answer to \textbf{Q3}.  Recall that we are using datasets in the source domain to obfuscate dataset in the target domain, and therefore, we can only manipulate dataset in the source domain.  Further, we select a few factors including size, diversity and similarity that may have impacts on the usability, and adjust the dataset using those factors accordingly to see what the impacts are. Assume that source domain consists of $n$ datasets which are denoted as $D_{source} = (d_1, d_2,...d_i,..., d_{n})$, where $d_{i}$ is referred to each dataset of the source domain. Particularly, we list those factors below: 
\begin{itemize}
\item \textbf{Size}.  We define the size of the source domain $D_{source} $ as follows:  
$$
    size(D_{source}) = \sum_{i=1}^{n} |d_{i}|
$$
\item \textbf{Diversity}. The diversity is defined as the amount of datasets involved in the source domain, which can be denoted as:
$$
    diversity(D_{source}) = |D_{source}|
$$
\item \textbf{Similarity}.  
To measure the similarity between domains, we first calculate the norm of a domain $\|D_{i}\|$ by calculating the average of all the samples in $D_{i}$, which can be regarded as a representative of the domain. Since we focus on image datasets, the representatives are essentially images. The similarity between domains is further defined as the similarity between the representatives of each domain.
To this end, we use perceptual hashing \cite{zauner2010implementation} to generate the fingerprints (a hash value) for each representative, denoted as $phash(\|D_{i}\|)$. By perceptual hashing,
if two images are perceptually identical, the difference between their fingerprints is modest \cite{zauner2010implementation}. Therefore, the similarity between domains is reflected by the difference between the fingerprints of their representatives. Formally, the similarity can be defined as follows:

\begin{footnotesize}
 
$$
    similarity = 1 - \frac{phash(\|D_{source}\|) - phash(\|D_{target}\|) }{len(phash(\|D_{source}\|))}
$$
 
\end{footnotesize}

\end{itemize}

In the following experiments, we select \texttt{Office-31} as the experiment dataset and discrepancy-based domain adaptions for DAMIA. Also, AlexNet is adopted as the backbone of the model.

\begin{figure}  
    \centering
    \includegraphics[width=0.45\textwidth]{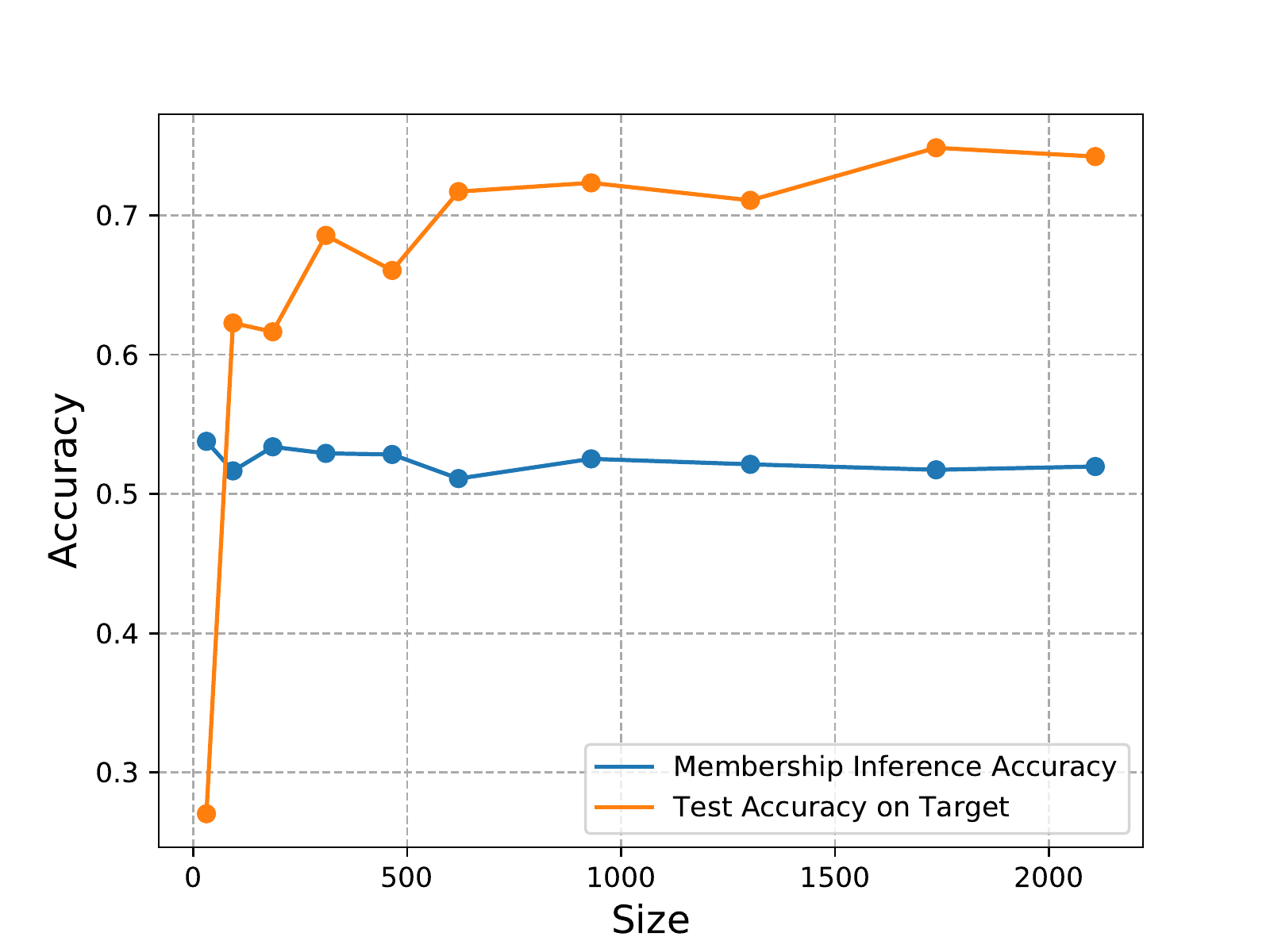}
    \caption{Test accuracy v.s. membership inference accuracy under different sizes of the dataset in source domain.  }
    \label{fig:exp03_size}
\end{figure}

\textbf{Size}: In our experiment, we use \texttt{Amazon} to obfuscate \texttt{Webcam}. Particularly, we want to explore how does the size of the dataset in the source domain (i.e., \texttt{Amazon}) affects the usability of the generated model. To this end, we gradually increase the size of each category in \texttt{Amazon} from 1 to 68. Recall that \texttt{Amazon} includes 31 categories in total, and therefore, the size of the source domain ranges from 31 to 2108. For each dataset with the specific size, we train the model for 100 epochs to minimize the impacts of errors. We then test the accuracy of the trained model. Provided that the obfuscated model can achieve a high accuracy, the usability of the model is not hindered. Moreover, changing the size of the dataset should not impair the effectiveness. Therefore, we also deploy the membership inference attacks against the trained model.  \autoref{fig:exp03_size} shows the result.  It can be observed that an increased size of a dataset in the source domain has negative impacts on the advantages of membership inference attacks slightly. This is reasonable since a larger dataset provides more samples for the model to perform the obfuscation. However, it can also be observed that even a dataset with a small size such as the one that has only 31 samples, also has excellent effectiveness. On the other hand, it can also be observed that a dataset with more samples has positive impacts on the usability.


\begin{table} 
    \centering
    \caption{Different sizes of dataset v.s. accuracy}
    \label{tab:exp03_size}
    \centering
    \resizebox{0.48\textwidth}{!}{%
    \begin{tabular}{cccccccc}
    \toprule
    \tabincell{c}{\textbf{Size of}\\ \textbf{Source Domain} }   & \tabincell{c}{\textbf{Train Acc.}\\ \textbf{on Target}}  &  \tabincell{c}{\textbf{Test Acc.}\\ \textbf{on Target}}  &  \tabincell{c}{\textbf{MIA Acc.}\\ \textbf{on Target}} & \tabincell{c}{\textbf{$adv_{mi}$}\\ \textbf{on Target}} \\ 
    \midrule
    \textbf{31  }  & 0.316037736   &  \textbf{0.270440252}   &  \textbf{0.537735849} & \textbf{0.075471698}      \\ 
    \rowcolor{table_gray}
    \textbf{93  }  & 0.555031447   &  \textbf{0.622641509}   &  \textbf{0.516509434} & \textbf{0.033018868}      \\ 
    \textbf{186 }  & 0.610062893   &  \textbf{0.616352201}   &  \textbf{0.533805031} & \textbf{0.067610062}      \\ 
    \rowcolor{table_gray}
    \textbf{310 }  & 0.682389937   &  \textbf{0.685534591}   &  \textbf{0.52908805 } & \textbf{0.0581761}        \\ 
    \textbf{465 }  & 0.694968553   &  \textbf{0.660377358}   &  \textbf{0.528301887} & \textbf{0.056603774}      \\ 
    \rowcolor{table_gray}
    \textbf{620 }  & 0.699685535   &  \textbf{0.716981132}   &  \textbf{0.511006289} & \textbf{0.022012578}      \\ 
    \textbf{930 }  & 0.724842767   &  \textbf{0.72327044 }   &  \textbf{0.525157233} & \textbf{0.050314466}      \\ 
    \rowcolor{table_gray}
    \textbf{1302}  & 0.746855346   &  \textbf{0.710691824}   &  \textbf{0.521226415} & \textbf{0.04245283}       \\ 
    \textbf{1736}  & 0.740566038   &  \textbf{0.748427673}   &  \textbf{0.517295597} & \textbf{0.034591194}      \\ 
    \rowcolor{table_gray}
    \textbf{2108}  & 0.732704403   &  \textbf{0.742138365}   &  \textbf{0.519654088} & \textbf{0.039308176}      \\ 
    \bottomrule
    \end{tabular}%
    } 
\end{table}

\textbf{Diversity}: We want to explore how does the diversity of the dataset in the source domain affects the usability of the model trained by DAMIA. Therefore, we modify the diversity of datasets in the source domain, and observe if the usability changes. Recall that there are three domains in \texttt{Office-31} including \texttt{Amazon}, \texttt{DSLR} and  \texttt{Webcam}.  We modify the diversity the datasets by ``mixing'' two of them together. For example,  \texttt{Mix(Amazon $\oplus$ DSLR)}  is referred to a new dataset that contains all the samples from the dataset \texttt{Amazon} and all samples from the dataset \texttt{DSLR}. 
Next, we use the mixed dataset to obfuscate the remaining one that is not involved in the mixing process. Similarly, all the models involved are trained for 100 epochs to minimize errors. We test the usability afterwards and deploy membership inference attacks against those models.  Results of diversity of 2 are illustrated in \autoref{tab:exp03_diversity}, and we compare the test accuracy under different diversities  (i.e., diversity equals 1 
\footnote{For source domains consisting of 1 dataset, their diversity is: $diversity(D_{source}) = |D_{source}| = |(d_1)| = 1$, such a source domain may be $D_{source} = (d_{Amazon})$. } 
and diversity equals 2 
\footnote{For source domains consisting of 2 datasets, their diversity is: $diversity(D_{source}) = |D_{source}| = |(d_1,d_2)| = 2$, such a  source domain may be $D_{source} = |(d_{Amazon},d_{DSLR})|   $.}) 
in \autoref{tab:exp03_diversity_target_test_acc} for better illustration. 
From the comparison, we see the diversity has a limited contribution to the usability. When diversity increases, the test accuracy does not always proportionally ascend. For example, the test accuracy of the model using \texttt{DSLR} to obfuscate \texttt{Webcam} is higher than the one using \texttt{Mix(Amazon $\oplus$ DSLR)} for obfuscation.  
Likewise, as shown in \autoref{tab:exp03_diversity_target_adv}, the comparison of the advantages of membership inference attacks with different diversities indicates that a larger diversity impairs the effectiveness of DAMIA. Therefore, DAMIA does not have advantages to be adopted to a dataset with a large diversity.



\begin{table}[H]
    \centering
    \caption{Test accuracy on mixed source domains}
    \label{tab:exp03_diversity}
    \centering
    \resizebox{0.48\textwidth}{!}{%
    \begin{tabular}{cccccccc}
    \toprule
    \tabincell{c}{\textbf{Adaptation} \\ \textbf{Direction}}   & \tabincell{c}{\textbf{Train Acc.}\\ \textbf{on Target}}  &  \tabincell{c}{\textbf{Test Acc.}\\ \textbf{on Target}}  &  \tabincell{c}{\textbf{MIA Acc.}\\ \textbf{on Target}} & \tabincell{c}{\textbf{$adv_{mi}$}\\ \textbf{on Target}} \\ 
    \midrule
    \textbf{Mix(DSLR $\oplus$ Webcam)} $\rightarrow$ \textbf{Amazon}  & 0.634265424   & \textbf{0.640070922}  & \textbf{0.501714027} & \textbf{0.003428054} \\ 
    \rowcolor{table_gray}
    \textbf{Mix(Amazon $\oplus$ Webcam)} $\rightarrow$ \textbf{DSLR  }  & 0.854271357   & \textbf{0.83       }  & \textbf{0.549346734} & \textbf{0.098693468} \\ 
    \textbf{Mix(Amazon $\oplus$ DSLR)} $\rightarrow$ \textbf{Webcam}  & 0.8525        & \textbf{0.8        }  & \textbf{0.56625    } & \textbf{0.1325}      \\ 
    \bottomrule
    \end{tabular}%
    } 
\end{table}

\begin{table*}
\centering
\caption{  Diversity of dataset in source domain v.s. accuracy. Please note that the diversity of ``Mix'' is set to 2.}
    \label{tab:exp03_diversity_target_test_acc}
    \begin{tabular}{ccccccc}
        \toprule
        Sensitive Dataset & \multicolumn{6}{c}{Dataset for Obfuscation}                                     \\
        \cmidrule{2-7}
              & Amazon & DSLR & Webcam & Mix(Amazon $\oplus$ DSLR) & Mix(Amazon $\oplus$ Webcam) & Mix(DSLR $\oplus$ Webcam) \\ 
        \midrule
        Amazon            & -           & 0.407900577 & 0.375055482 & -                & -                  & \textbf{0.64007092} \\
        \rowcolor{table_gray}
        DSLR              & 0.46        & -           & 0.929648241 & -                & \textbf{0.83}      & -                   \\
        Webcam            & 0.459119497 & 0.938679245 & -           & \textbf{0.80}    & -                  & -                   \\
        \bottomrule
    \end{tabular}
\end{table*}


\begin{table*}
\centering
\caption{Diversity vs advantages. Please note that the diversity of ``Mix'' is set to 2.}
\label{tab:exp03_diversity_target_adv}
    \begin{tabular}{ccccccc}
        \toprule
        Sensitive Dataset & \multicolumn{6}{c}{Dataset for Obfuscation}                                     \\
        \cmidrule{2-7}
              & Amazon & DSLR & Webcam & Mix(Amazon $\oplus$ DSLR) & Mix(Amazon $\oplus$ Webcam) & Mix(DSLR $\oplus$ Webcam) \\ 
        \midrule
        Amazon            & -           & 0.015031966 & 0.003106968 & -                & -                  & \textbf{0.003428054} \\
        \rowcolor{table_gray}
        DSLR              & 0.085929648        & -           & 0.103819096 & -                & \textbf{0.098693468}      & -                   \\
        Webcam            & 0.034591194 & 0.009433962 & -           & \textbf{0.1325}    & -                  & -                   \\
        \bottomrule
    \end{tabular}
\end{table*}

\begin{figure*} 
    \centering
    \includegraphics[width=0.85\textwidth]{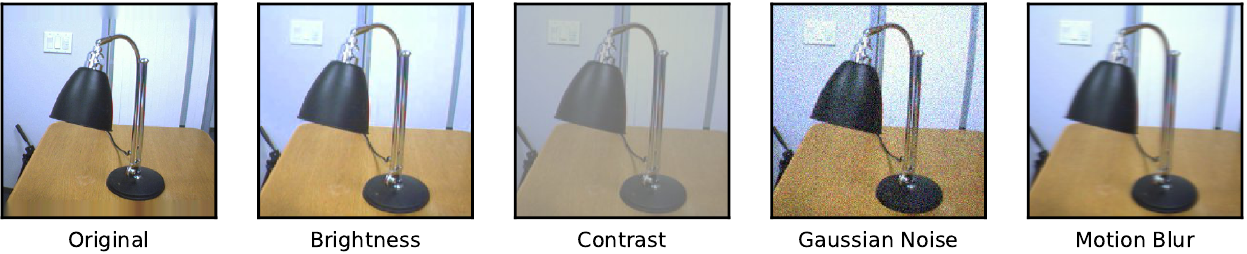}
    \caption{Images of a desk lamp from \texttt{Webcam} and its perturbed versions.  }
    \label{fig:exp03_similarity}
\end{figure*}

\textbf{Similarity}:  We then explore that how does the similarity between the source and target domains affect the usability of DAMIA. To this end, we create multiple similar datasets to obfuscate \texttt{Webcam}. Instead of collecting similar samples, we generate similar samples by modifying original samples from \texttt{Webcam} slightly. For example, the modification can be realized through the adjustment of the luminance (brightness) and tools for that goal  \cite{hendrycks2019benchmarking} are available online \footnote{https://github.com/hendrycks/robustness}. Particularly,  brightness, contrast, Gaussian noise and motion blur are involved in our experiment to modify the original image as shown in  \autoref{fig:exp03_similarity}. Similar to the previous experiments, all the models are trained for 100 epochs, and membership inference attacks are deployed on those models. \autoref{tab:exp03_similarity} shows that the test accuracies of models are specifically enhanced with an excellent effectiveness achieved.

\begin{table}[H]
    \centering
    \caption{Similarity v.s. Test accuracy. The \texttt{Original} stands for the \texttt{Webcam} from \texttt{Office-31}. Please note that in this case, the similarity between the similar source and the target is above 0.8, while in the case where the sources are \texttt{Amazon} and \texttt{DSLR}, the similarities are \textbf{0.593750} and \textbf{0.781250}.}
    \label{tab:exp03_similarity}
    \centering
    \resizebox{0.48\textwidth}{!}{%
    \begin{tabular}{cccccccc}
    \toprule
    \tabincell{c}{\textbf{Adaptation} \\ \textbf{Direction}} & \tabincell{c}{ \textbf{Similarity} } & \tabincell{c}{\textbf{Train Acc.}\\ \textbf{on Target}}  &  \tabincell{c}{\textbf{Test Acc.}\\ \textbf{on Target}}  &  \tabincell{c}{\textbf{MIA Acc.}\\ \textbf{on Target}} & \tabincell{c}{\textbf{$adv_{mi}$}\\ \textbf{on Target}} \\ 
    \midrule
    \textbf{Brightness $\rightarrow$  Original   }  &  0.843750  & 0.993710692    & \textbf{1.0}         & \textbf{0.510220126} & \textbf{0.020440252}\\ 
    \rowcolor{table_gray}
    \textbf{Contrast $\rightarrow$  Original     }  &  0.843750  & 0.993710692    & \textbf{0.993710692} & \textbf{0.512578616} & \textbf{0.025157232}\\ 
    \textbf{Gaussian Noise $\rightarrow$ Original}  &  0.812500  & 0.988993711    & \textbf{0.974842767} & \textbf{0.522012579} & \textbf{0.044025158}\\ 
    \rowcolor{table_gray}
    \textbf{Motion Blur $\rightarrow$  Original  }  &  0.812500  & 0.995283019    & \textbf{0.993710692} & \textbf{0.512578616} & \textbf{0.025157232}\\ 
    \bottomrule
    \end{tabular}%
    } 
\end{table}

\section{Related Works} 
\label{sec:related_works}

\textbf{Defenses against Membership Inference Attacks}: 
Defenses against membership inference can be categorized into three groups: (i) regularization-based defenses; (ii) differential-privacy-based defenses; (iii) adversarial-attack-based defenses. Regularization-based defenses directly adopt regularization techniques to build defenses. \citet{shokri2017membership} and \citet{salem2018ml} show that potential techniques including $L_2$ regularization \cite{ng2004feature} and Dropout \cite{srivastava2014dropout} may prevent overfitting issues so as to counter the membership inference attacks. However, our defense is stemmed from the DA techniques, which are not designed to address the overfitting issues.  \citet{salem2018ml} use another regularization technique called ensemble learning to build their defense. Their defense requires extra storage to maintain ML (i.e. Machine Learning) models. Our defense does not have such a requirement. \citet{nasr2018machine} introduce a new regularization term and propose an adversarial training process termed min-max game to optimized the regularization terms so as to defend against the membership inference attacks. However, their defense is time-consuming since the adversarial training process is involved. Adversarial-attack-based defenses shield victim models by adversarial attacks. \citet{jia2019memguard} propose MemGuard, where adversarial examples \cite{goodfellow2014explaining} are introduced to obfuscate samples and confuse the attackers. However, MemGuard is subject to sophisticated operations to turn the outputs of victim models into adversarial examples. When compared with MemGuard, our defense can achieve the same goal while requiring fewer efforts.  
Differential-privacy based defenses have draw defenders' attention \cite{pyrgelis2017knock, triastcyn2018generating, xu2019ganobfuscator, jayaraman2019evaluating}, where the goal is achieved via adding noises to the loss function or gradients of the model. However, the method also degrades the usability of the model and slows the training process. DAMIA does not add noises to the loss function, and the efficiency is not hindered while training when compared with their scheme.

\textbf{Transfer Learning in Privacy-Preserving Machine Learning.} \citet{papernot2016semi} apply transfer learning to avoid model information leakage.  \citet{triastcyn2018generating} adopt GAN \cite{goodfellow2014generative} to address a similar issue as the work \cite{papernot2016semi}. Particularly, in \cite{triastcyn2018generating}, the authors generate the artificial training dataset based on a private dataset and use the generated dataset to train a model. However, those efforts do not have a focus, stating what type of attack their defense may counter. Our DAMIA focuses on countering membership inference attacks.  \citet{shokri2017membership,song2019privacy} shows that the temperature scaling technique, which is widely adopted in the area of domain adaptation, has the potential to defend membership inference attacks. In our defense, we use the other domain adaptation techniques rather than the temperature scaling technique to defend against the attacks.

\section{Conclusion} 
\label{sec:conclusion}

 In this paper, we propose DAMIA, which leverages domain adaptation as a defense to prevent membership inference attacks. DAMIA effectively counters the membership inference attacks while the usability is not hindered. Also, we show that with proper factors enabled, the performance of the model can be boosted. The next stage of our work is to explore a mechanism that can automatically select or generate a related dataset of the given sensitive dataset so as to free the manual load.

\section*{Acknowledgements}


Weiqi Luo was partially supported by National Natural Science Foundation of China (Grant No. 61877029), Guangdong Provincial Special Funds for Applied Technology Research and Development and Transformation of Important Scientific and Technological Achieve (Grant Nos.2017B010124002). 
Jian Weng was partially supported by National Key R\&D Plan of China (Grant Nos. 2017YFB0802203, 2018YFB1003701), National Natural Science Foundation of China (Grant Nos. 61825203, U1736203, 61732021), Guangdong Provincial Special Funds for Applied Technology Research and Development and Transformation of Important Scientific and Technological Achieve (Grant Nos. 2016B010124009 and 2017B010124002). 
Guoqiang Zeng was partially supported by National Natural Science Foundation of China (Grant Nos. 11871248, U1636209). 
Yue Zhang was partially supported by National Natural Science Foundation of China (Grant Nos. 61877029). 
Hongwei Huang was partially supported by National Natural Science Foundation of China (Grant Nos. 61872153). 
Anjia Yang was partially supported by National Natural Science Foundation of China (Grant Nos. 61702222).


\bibliographystyle{IEEEtranN}
\bibliography{references}

\begin{IEEEbiography}[{\includegraphics[width=1in,height=1.25in,clip,keepaspectratio]{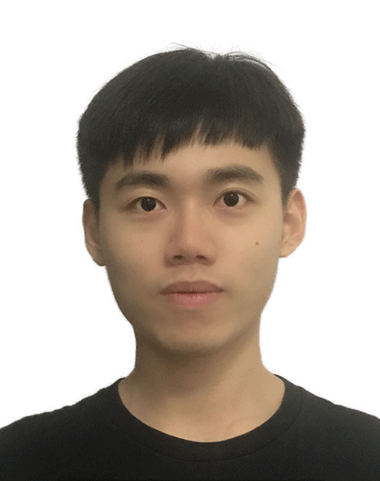}}]{Hongwei Huang} 
received his B.S. degree in software engineering from South China Agriculture University (2014 - 2018). He has been pursuing his M.S. degree in Jinan University since 2018. His research interests include machine learning and its privacy and security.
\end{IEEEbiography}

\begin{IEEEbiography}[{\includegraphics[width=1in,height=1.25in,clip,keepaspectratio]{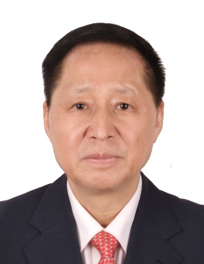}}]{Weiqi Luo} received his B.S. degree from Jinan University in 1982 and Ph.D. degree from South China University of Technology in 2000. Currently, he is a professor with School of Information Science and Technology in Jinan University, Guangzhou. His research interests include network security, big data, artificial intelligence, etc. He has published more than 100 high-quality papers in international journals and conferences.
\end{IEEEbiography}

\begin{IEEEbiography}[{\includegraphics[width=1in,height=1.25in,clip,keepaspectratio]{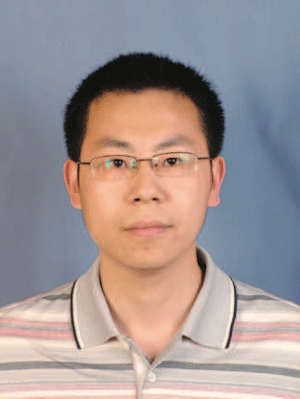}}]{Guoqiang Zeng} received Ph.D. degree in Control Science and Engineering from Zhejiang University, China, in 2011. He is currently a professor in College of Cyber Security, Jinan University, Guangzhou, and is also with National-Local Joint Engineering Laboratory of Digitalize Electrical Design Technology, Wenzhou University, Wenzhou, China. His research interests include modeling, control and optimization of smart grid, Internet of Things, computational intelligence and cyber security. He has authored or co-authored the book Extremal optimization: Fundamentals, algorithms, and applications published by CRC press, and more than 50 journal and conference articles. He holds over ten patents. Dr. Zeng was a recipient of eight ministerial and provincial science and technology progress awards.
\end{IEEEbiography}

\begin{IEEEbiography}[{\includegraphics[width=1in,height=1.25in,clip,keepaspectratio]{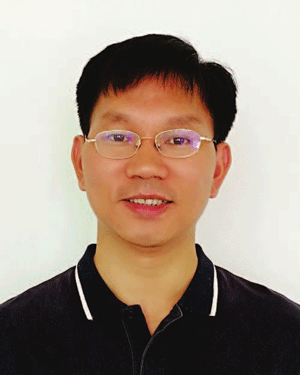}}]{Jian Weng} is a professor and the Executive Dean with College of Information Science and Technology in Jinan University. He received B.S. degree and M.S. degree at South China University of Technology in 2001 and 2004 respectively, and Ph.D. degree at Shanghai Jiao Tong University in 2008. His research areas include public key cryptography, cloud security, blockchain, etc. He has published 80 papers in international conferences and journals such as CRYPTO, EUROCRYPT, ASIACRYPT, TCC, PKC, CT-RSA, IEEE TDSC, etc. He also serves as associate editor of IEEE Transactions on Vehicular Technology.
\end{IEEEbiography}

\begin{IEEEbiography}[{\includegraphics[width=1in,height=1.25in,clip,keepaspectratio]{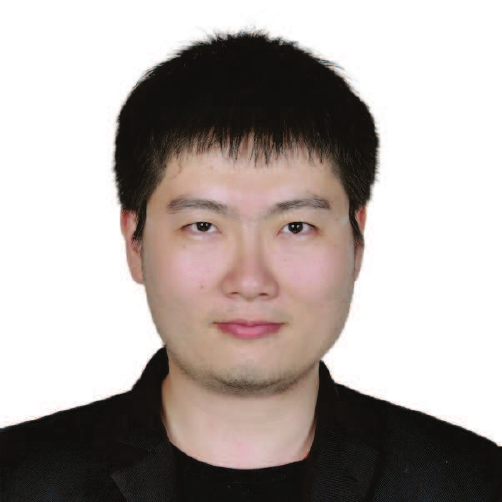}}]{Yue Zhang}  is a PhD student in the College of Information Science and Technology \& College of Cyber Security at Jinan University. Also, he studied and worked at the University of Central Florida (UCF) / University of Massachusetts Lowell (UML). His research focuses on system security, especially IoT security. He won Outstanding Research Paper Award of GuangDong Computer Federation in 2019, Best Paper Award of IEEE International Conference on Industrial Internet in 2019, National scholarship for Ph.D Students in Cyber Security (China),  Outstanding Graduates in 2016, National scholarship for Master Students in 2015. He has been publishing papers in conferences such as IEEE International Conference on Computer Communications (INFOCOM), International Symposium on Research in Attacks, Intrusions and Defenses (RAID) journals such as IEEE Transactions on Dependable and Secure Computing (TDSC), IEEE Transactions on Parallel and Distributed Systems (TPDS). His research work was also selected by Black Hat Aisa.
\end{IEEEbiography}

\begin{IEEEbiography}[{\includegraphics[width=1in,height=1.25in,clip,keepaspectratio]{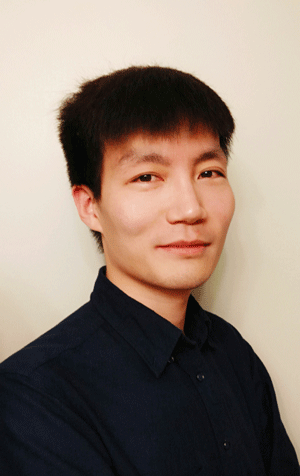}}]{Anjia Yang}   (M'17) received the B.S. degree from Jilin University in 2011 and the Ph.D. degree from the City University of Hong Kong in 2015. From December 2015 to September 2019, he held postdoc positions in City University of Hong Kong and Jinan University, respectively, during when he visited University of Waterloo in Canada. He is currently an associate professor in Jinan University, Guangzhou. His research interests include security and privacy in vehicular networks, internet of things,  blockchain and cloud computing, and he has published more than 20 papers.
\end{IEEEbiography}

\end{document}